\documentclass[trackchanges]{aastex7}
\usepackage{amsmath}
\usepackage[T1]{fontenc}
\usepackage[normalem]{ulem}
\usepackage{hyperref}   
\usepackage[toc,page]{appendix} 

\begin{document}

\title{ Light curves of time-dependent accretion disk in tidal disruption events }

\author[]{Chenlei Guo}
\affiliation{National Astronomical Observatories, Chinese Academy of Sciences, Beijing 100101, China; guocl@bao.ac.cn, qiaoel@nao.cas.cn}
\affiliation{School of Astronomy and Space Sciences, University of Chinese Academy of Sciences, 19A Yuquan Road, Beijing 100049, China}
\email{guocl@bao.ac.cn}

\author[]{Erlin Qiao\thanks{Corresponding author}}
\affiliation{National Astronomical Observatories, Chinese Academy of Sciences, Beijing 100101, China; guocl@bao.ac.cn, qiaoel@nao.cas.cn}
\affiliation{School of Astronomy and Space Sciences, University of Chinese Academy of Sciences, 19A Yuquan Road, Beijing 100049, China}
\email{qiaoel@nao.cas.cn}

\correspondingauthor{Erlin Qiao}
\email[show]{qiaoel@nao.cas.cn}


\begin{abstract}
Tidal disruption events (TDEs) are believed to be an ideal laboratory for studying the evolution of accretion flow around a supermassive black hole (BH). In general, the mass feeding rate to the BH is suggested to be super-Eddington initially, and evolves to be sub-Eddington on timescales of years. In this paper, we carry out calculations of the time-dependent evolution of accretion disk in the standard environment of TDE, 
i.e., injecting matter at the circularization radius of the stellar debris in the form of $\dot M_{\rm inject} \propto t^{-5/3}$. We find that when $\dot M_{\rm inject}$ evolves to a value around the Eddington accretion rate, the radiation pressure instability occurs. 
We test the influence of the model parameters on the light curves, such as the BH mass $M_{\rm BH}$, viscosity parameter $\alpha$, and 
 mass-injecting radius $R_{\rm{out}}$, 
all of which are found to affect the light curves to some extent. In most cases, we find that the light curves oscillate significantly due to the radiation pressure instability. As an exception, when $\alpha$ is small or $R_{\rm{out}}$ is large, we find that the oscillations are completely suppressed. In this case, the light curve drops steeply and then becomes flat in the late-time evolution, which we apply to explain the observed ultraviolet (UV) light curves of ASASSN-15oi and ASASSN-14ae together with the assumption of a photosphere.
Finally, we discuss the potential applications of our time-dependent accretion disk model to explaining multi-band light curves of TDEs in the future.
\end{abstract}



\section{Introduction}
{
A tidal disruption event (TDE) occurs when a star moves close enough to a supermassive black hole (BH). The star is torn apart when the tidal force of the BH exceeds its self-gravity. This happens at the tidal disruption radius, $R_{\rm{T}}=R_{\star}(M_{\rm{BH}}/M_{\star})^{1/3}$ for a BH with mass $M_{\rm{BH}}$, a star with mass $M_{\star}$ and radius $R_{\star}$. 
About half of the stellar debris escapes from the BH while the rest of them remain bound and fall back to the BH.
The fallback rate is derived to be super-Eddington initially and decays in the form of $t^{-5/3}$ (\cite{Rees1988}, \cite{Phinney1989}). 
These fallback debris generate the main radiations observed in TDEs through various processes, including collisions, accretion, reprocessing, etc. 
In general, the X-ray emissions in TDEs are mostly thought to be produced by an accretion disk. The origin of the optical/UV emission is not determined, but could be powered by the shock induced by the collision of the debris (\cite{Piran_2015}, \cite{Jiang2016ApJ}, \cite{Bonnerot_2017MNRAS}, \cite{Steinberg_2024Natur}), or could be explained by the ‘reprocessing’
model with the X-rays and extreme ultraviolet (EUV) radiation reprocessed into optical/UV band by a surrounding optically thick envelope or outflow (\cite{Ulmer1998A&A}, \cite{Ulmer1999ApJ}, \cite{Metzger_Stone2016MNRAS}, \cite{Dai2018ApJ}, \cite{Qiao2025MNRAS.539.3473Q}). 

As a powerful source of energy, accretion is critical for producing the diverse radiation observed in TDEs (\cite{Lodato2011MNRAS}, \cite{Shen2014ApJ}, \cite{Piro_2025}, \cite{Mummery_Balbus2020MNRAS}, \cite{Alush_2025ApJ_1}). Because of 
the relative quick evolution of the accretion flow, 
the study of the time-dependent accretion disk is particularly necessary in TDEs. 
\cite{Shen2014ApJ} proposed a one-zone (zero-dimensional) time-dependent disk model for studying the temporal evolution of TDEs. 
Then a more detailed study based on it is carried out by \cite{Piro_2025}. In these models, the disk is treated as a uniform spreading torus, with continuous mass injection in the form of the fallback rate. One of the main findings of these works is that the radiation pressure instability occurs when the fallback rate decays to certain values near the Eddington accretion rate.

Another approach to studying the time-dependent disk in TDEs is the one-dimensional (1D) spreading thin disk models (\cite{Cannizzo_1990ApJ}, \cite{Velzen2019ApJ}, \cite{Mummery_Balbus2020MNRAS}, \cite{Alush_2025ApJ_1}). These models assume the disk to be geometrically thin and freely expanding. The radiation pressure instability is fully suppressed by considering either different descriptions for viscosity (\cite{Mummery_Balbus2020MNRAS}) or strongly magnetized disk(\cite{Alush_2025ApJ_1}).
These models have successfully explained the optical/UV plateaus observed in the late times of TDEs (\cite{Mummery2024MNRAS},  \cite{Mummery2025MNRAS.541..429M_Ref6}, \cite{Guolo2025arXiv251026774G_Ref2}, \cite{2025arXiv251024696A_Ref3}, \cite{2025ApJ...992..114G_Ref4}). 

For most TDEs, the peak of the fallback rate is typically 1-2 orders of magnitude higher than the Eddington accretion rate $\dot{M}_{\rm{Edd}}$ (defined as $L_{\rm{Edd}}/(0.1 c^{2})$, where $L_{\rm{Edd}}=1.26\times 10^{38}M_{\rm{BH}}/M_{\odot}\rm{erg/s}$  is the Eddington luminosity). If the debris can efficiently circularize to form a disk, the disk should initially be a geometrically thick, radiation pressure dominated slim disk (\cite{Dai2018ApJ}, \cite{Qiao2025MNRAS.539.3473Q}). As the fallback rate decays to be sub-Eddington, we expect to witness the transition of a slim disk to a thin disk, during which the radiation pressure instability  is suggested to occur in the standard framework of accretion disk (\cite{Shakura_Sunyaev1976MNRAS}, \cite{LE1974}, \cite{Abramowicz1988}). To investigate whether and how this instability occurs in TDEs, more detailed calculations and comparisons with observations are required. 

In this work, we perform calculations of the time-dependent evolution of the accretion disk in standard TDE environment, i.e., injecting matter at the circularization radius of the stellar debris in the form of $\dot{M}_{\rm{inject}}\propto t^{-5/3}$.  
We find that the radiation pressure instability occurs when $\dot M_{\rm inject}$ evolves to a value around $\dot M_{\rm Edd}$, and that $M_{\rm BH}$, viscosity parameter $\alpha$, and mass-injecting radius $R_{\rm{out}}$ all show effects on the light curve to some extent. In most cases, the radiation pressure instability causes the light curves to oscillate significantly. As an exception, the oscillations are completely suppressed when $\alpha$ is small or $R_{\rm{out}}$ is large. In this case, the light curve drops steeply and then becomes flat in the late times. We apply this kind of light curves to fit the observed UV light curves of ASASSN-15oi and ASASSN-14ae together with the assumption of a photosphere. It is found that the photosphere radius is initially much larger than the size of the disk, indicating strong reprocessing in the early times. While in the late times, the photosphere radius becomes comparable to the disk outer radius, and the UV flux from the photosphere model is similar to that of the disk model.  
We demonstrate with these two examples that some steep decay in the observed UV light curves could possibly be attributed to the radiation pressure instability. 

This paper is organized as follows. In Section 2, we present our disk model and derive the fallback rate in TDEs. In Section 3, we display our results and test the effect of different parameters on the light curve. In Section 4, we apply our disk model together with a photosphere to fit the UV emissions of ASASSN-15oi and ASASSN-14ae. In Section 5, we compare our results with previous works, and discuss the improvement of our model in future work and other potential applications to observations. Summary and conclusions are presented in Section 6.

\section{Model}
\subsection{Basic equations of time-dependent accretion disk}
\label{Disk model}
Below we list the equations that describe a time-dependent accretion disk. The mass conservation equation is,
\begin{equation}
    \frac{\partial \Sigma}{\partial t} = \frac{1}{2\pi r}\frac{\partial \dot{M}}{\partial r}, \label{mass conserv}
\end{equation}
where
\begin{equation}
    \dot{M} = -2\pi r\Sigma v_{\rm{r}}. \label{M_dot}
\end{equation}
The angular momentum conservation equation is,
\begin{equation}
    \frac{\partial }{\partial t}\left (\Sigma r^{2}\Omega \right ) = \frac{1}{2\pi r}\frac{\partial}{\partial r}\left (\dot{M}r^{2}\Omega \right ) + \frac{1}{2\pi r}\frac{\partial}{\partial r}(2\pi r^{2}T_{\rm{r\varphi}}). \label{angular momentum conserv}
\end{equation}
The energy conservation equation is,
\begin{equation}
    \Sigma T \frac{d S}{dt}=Q^{+}_{\rm{vis}} - Q^{-}_{\rm{rad}} .\label{energy conserv1}
\end{equation}
The equations (\ref{mass conserv}), (\ref{M_dot}), (\ref{angular momentum conserv}) and (\ref{energy conserv1}) are vertically integrated over the disk. $\Sigma$ is the surface density, $T$ is the temperature, $\Omega$ is the angular velocity, $v_{\rm{r}}$ is the radial velocity, and $S$ is the entropy. In equation \ref{energy conserv1}, the term $\Sigma T \frac{dS}{dt}$ on the left hand side represents the advection of energy, $Q^{+}_{\rm{vis}}$ is the viscous heating rate and $Q^{-}_{\rm{rad}}$ is the radiative cooling rate per unit surface area.

We adopt the Keplerian angular velocity in a Newtonian potential, so the angular velocity $\Omega=\Omega_{\rm{K}}=\sqrt{GM_{\rm BH}/ r^3}$, where $G$ is the gravitational constant and $M_{\rm BH}$ is the BH mass. The vertically integrated $r\varphi$ component of the stress tensor is,
\begin{equation}
    T_{\rm{r\varphi}}=\int_{-H}^{H} -\alpha p dz=-2C_{1}\alpha p_{\rm{e}}H, \label{Trphi}
\end{equation}
where $\alpha$ is the dimensionless viscosity parameter, $H$ is the half-thickness of the disk, $p_{\rm{e}}$ is the total pressure on the equatorial plane of the disk, and $C_{1}$ (also $C_{2}$ and $C_{3}$ in the following equations) is a vertical integration coefficient, described in detail in Appendix \ref{app:vertical}.
The total pressure $p$ in the disk can be expressed as,
\begin{equation}
    p = \frac{k}{\mu m_{\rm{p}}}\rho T + \frac{1}{3}aT^{4}. \label{total pressure}
\end{equation}
where $\rho$ is the density of the accretion flow. We take the mean molecular weight to be $\mu=0.6$ assuming the solar metallicity. 

The hydrostatic equilibrium in the vertical direction can be written as, 
\begin{equation}
    \frac{p_{\rm{e}}}{\rho_{\rm{e}}} = C_{3}\Omega_{\rm{K}}^{2}H^{2}, \label{hydrostatic equilibrium}
\end{equation}
where $\rho_{\rm{e}}$ is the density on the equatorial plane of the disk.
Combining equation (\ref{mass conserv}), (\ref{M_dot}) and (\ref{angular momentum conserv}), together with (\ref{Trphi}) and (\ref{hydrostatic equilibrium}),  we obtain the expression for the radial velocity $v_{\rm{r}}$ as \footnote{In deriving this equation, we adopt an alternative expression for $T_{\rm{r\varphi}}$ in terms of $\nu$ and $\Sigma$, i.e., $T_{\rm{r\varphi}}=-2C_{1}\alpha p_{\rm{e}}H=-2C_{1}\alpha \rho_{\rm{e}} c_{\rm{s}}^{2}H
=-2C_{1}\sqrt{C_{3}}\frac{3}{2}\nu \Sigma \Omega_{\rm{K}} $. }
\begin{equation}
    v_{\rm{r}} = -\frac{6C_{1}\sqrt{C_{3}}}{\Sigma r^{1/2}}\frac{\partial}{\partial r}(\nu \Sigma r^{1/2}). \label{vr}
\end{equation}
where $\nu$ is the kinematic viscosity. We have taken the standard $\alpha$ prescription for viscosity, i.e., $\nu = (2/3)\alpha H c_{\rm{s}}$, where $c_{\rm{s}}$ is the isothermal sound speed defined as $c_{\rm s}=\sqrt{p_{\rm{e}}/\rho_{\rm{e}}}$.

The viscous heating rate per unit surface area (two sides) is, 
\begin{equation}
  Q^{+}_{\rm{vis}}=rT_{\rm{r\varphi}}\frac{d\Omega}{d \mathit{r}}= 2C_{\rm{1}}\frac{3}{2}\alpha p_{\rm{e}}H\mathit{\Omega}. \label{Q_vis}
\end{equation} 
The radiative cooling rate per unit surface area is,
\begin{equation}
     Q^{-}_{\rm{rad}} = 2 C_{2}\frac{4\sigma T^{4}}{3\kappa \Sigma}, \label{Qrad}
\end{equation}
where $\sigma$ is the Stefan-Boltzmann constant, and $\kappa$ is the electron scattering opacity with $\kappa=0.34 \,\rm{cm^{2}g^{-1}}$ adopted.

In order to calculate the entropy gradient, we use the first law of thermodynamics,
\begin{equation}
     TdS=de + pd(1/\rho),
\end{equation}
where $e$ is the internal energy per unit mass.  
The internal energy per unit volume is $\rho e = 3/2p_{\rm{gas}} + 3p_{\rm{rad}}$ with adiabatic index $\gamma=5/3$. The 
vertically integrated internal energy can be written as\footnote{Here $\int_{-H}^{H}aT^{4}dz=2C_{1}aT_{\rm{e}}^{4}H \approx aT_{\rm{e}}^{4}H $ for $C_{1}=0.46$ (see details in Appendix \ref{app:vertical}), which is similar to $\int_{-H}^{H}pdz$. $T_{\rm{e}}$ is the temperature on the equatorial plane. We use the disk temperature $T$ as an approximation for $T_{\rm{e}}$ .}
\begin{equation}
    E = \frac{3}{2}\frac{k}{\mu m_{\rm{p}}}\Sigma T + aT^{4}H. \label{internal energy E}
\end{equation}

The energy equation (\ref{energy conserv1}) can be then expressed as, 
\begin{equation}
    \frac{\partial E}{\partial t} + \frac{\partial(rE v_{\rm{r}})}{r\partial r} + \Pi\frac{\partial(r v_{\rm{r}})}{r\partial r}= Q^{+}_{\rm{vis}} - Q^{-}_{\rm{rad}}, \label{energy conserv2}
\end{equation} 
where $\Pi=2C_{1}p_{\rm{e}}H$. 

In summary, combining equation (\ref{total pressure}), (\ref{hydrostatic equilibrium}),  (\ref{vr}) and (\ref{internal energy E}), we can solve equation  (\ref {mass conserv}) and (\ref {energy conserv2}) by specifying BH mass $M_{\rm{BH}}$, viscosity parameter $\alpha$, the mass injecting rate $\dot{M}_{\rm inject}$, and the radius $R_{\rm{out}}$ where the mass is injected. In this paper, equation (\ref {mass conserv}) and (\ref {energy conserv2}) are solved by the implicit finite volume method. We set the inner boundary at the innermost stable circular orbit $R_{\rm{ISCO}}=3R_{\rm{S}}$ for a non-rotating BH, where $R_{\rm{S}}$ is the Schwarzschild radius defined as $R_{\rm{S}}\equiv 2GM_{\rm{BH}}/c^{2}$. We adopt $\Sigma_{\rm{in}}=E_{\rm{in}}=0$ for the inner boundary condition. The outer boundary condition of $\Sigma_{\rm{out}}$ and $E_{\rm{out}}$ is calculated by a steady disk (see details in Appendix \ref{app:steady solutions}) with a given mass injection rate, as in that of \cite{Janiuk2002ApJ}. 
We discuss this kind of setting of outer boundary condition in Section \ref{Discuss_Comparison}.

\subsection{Fallback rate in TDEs}
We assume that the initial orbit of the star is parabolic with a pericenter distance $R_{\rm{p}}$. The penetration factor of the star is defined as $\beta=R_{\rm{T}}/R_{\rm{p}}$.
We consider the case of $\beta=1$. 
We follow the simple assumption that the star is almost unperturbed until it reaches pericenter, where an impulsive interaction is exerted by the BH and the star is torn apart \cite[][]{Rees1988}.
In this approximation, the energy spread among the debris of the disrupted star in the BH gravitational potential can be estimated as,
\begin{equation}
    \Delta E = \frac{GM_{\rm{BH}}}{R_{\rm{p}}^{2}}R_{\star}. \label{E_spread}
\end{equation}
In this case, half of the debris will escape away from the BH, while the remaining half of the debris will return to the BH assuming to follow a Keplerian trajectory.  
The return time $T$ of the debris with specific orbital energy $E_{\rm{orb}}$ can be expressed as,
\begin{equation}
    T=\frac{2\pi GM_{\rm{BH}}}{(2|E_{\rm{orb}}|)^{3/2}}.
\end{equation}
Taking $E_{\rm{orb}}=-\Delta E$ for the most bound debris, we obtain the fallback timescale
\begin{equation}
    t_{\rm{fb}} \approx 40M_{6}^{1/2}m_{\star}^{-1}r_{\star}^{3/2} \rm{days}, \label{tfb}
\end{equation}
where $M_{6}=M_{\rm{BH}}/(10^{6}M_{\odot})$, $m_{\star}=M_{\star}/M_{\odot}$ and $r_{\star}=R_{\star}/R_{\odot}$.

The fallback rate can be expressed as, 
\begin{equation}
   \dot{M}_{\rm{fb}}=\frac{dM}{dt}=\frac{dM}{dE}\frac{dE}{dt}=\frac{M_{\star}}{3t_{\rm{fb}}} \left ( \frac{t}{t_{\rm{fb}}} \right )^{-5/3}, \label{M_dot_fb}
\end{equation}
where we have adopted $dM/dE=M_{\star}/(2\Delta E)={\rm constant}$, i.e., assuming a flat orbital energy distribution among the bound debris, and $dE/dt=d(E_{\rm{orb}})/dT$. The peak fallback rate is
\begin{equation}
    \dot{M}_{\rm{peak}}=\frac{M_{\star}}{3t_{\rm{fb}}}\approx 133.8\dot{M}_{\rm{Edd}}  M_{6}^{-3/2}m_{\star}^{2} r_{\star}^{-3/2}. \label{M_dot_peak}
\end{equation}

The fallback debris will be circularized through collisions or the physics we don't know clearly 
currently. According to angular momentum conservation, the circularization radius is calculated 
to be
\begin{equation}
    R_{\rm{c}}=2R_{\rm{p}}=1.4\times 10^{13}M_{6}^{1/3}m_{\star}^{-1/3} r_{\star}\rm{cm}. \label{Rc}
\end{equation}
In this paper, we calculate the structure and the evolution of the accretion disk
considering the TDE environment, i.e., continuously injecting mass at the circularization radius $R_{\rm{c}}$ in the form of the fallback rate (equation (\ref {M_dot_fb})) as the mass supply to the BH.

\section{Results}
\label{Results_section}
Disk luminosity and the corresponding light curve are key physical quantities, which
can be connected to observations. We calculate the disk luminosity by integrating the emission of each disk annulus
from ISCO to $R_{\rm out}$ as follows, 
\begin{equation}
    L=2\int_{R_{\rm{ISCO}}}^{R_{\rm{out}}}\sigma T_{\rm{eff}}^{4} 2\pi rdr.
    \label{Luminosity}
\end{equation}
Here $T_{\rm{eff}}$ is the effective temperate of the accretion disk at some fixed radius, and can be calculated with the formula of $T_{\rm{eff}} =[{Q^{-}_{\rm{rad}}/2\sigma}]^{1/4}$. In this paper, by solving the
time-dependent accretion disk in standard TDE environment, we can work out the disk luminosity and the 
corresponding light curve.

In the upper panel of Figure \ref{Lum_Scurve}, we plot 
a standard light curve from our calculations by setting $M_{\rm{BH}}=10^{6}M_{\odot}$, $M_{\star}=M_{\odot}$ and $R_{\star}=R_{\odot}$ (unless stated otherwise, we adopt $M_{\star}=M_{\odot}$ and $R_{\star}=R_{\odot}$ throughout the paper), $\alpha=0.01$. The mass debris is set to be injected at $R_{\rm out}=R_{\rm c}$ in the form of $\dot M_{\rm inject}=\dot M_{\rm fb}\propto t^{-5/3} $ as equation (\ref {M_dot_fb}). 
The peak value of fallback rate $\dot M_{\rm peak}$ is calculated to be $133.8\dot M_{\rm Edd}$, and 
$R_{\rm c}$ is calculated to be $47R_{\rm S}$ according to equation (\ref{M_dot_peak}) and (\ref{Rc}) respectively. Here $R_{\rm{S}}$ is the Schwarzschild radius defined as $R_{\rm{S}}\equiv 2GM_{\rm{BH}}/c^{2}$.
In order to more clearly show the evolution of the light curve at different times,
we plot the light curve with different colors for different times. One can see the colorful light curve in the upper panel of Figure \ref {Lum_Scurve} for details. As a comparison, we also plot the mass injection rate 
(converted to luminosity by assuming $L=0.1\dot{M}_{\rm{inject}}c^{2}$)
as a function of time.  One can see the black dashed line in the upper panel of Figure \ref {Lum_Scurve} for details. 
The corresponding evolution of the accretion disk in $T_{\rm eff}-\Sigma$ diagram at some fixed radius is plotted 
in the lower panel of Figure \ref{Lum_Scurve}, in which the so-called S-shaped curve (black solid line) is also added for comparison. The S-shaped curve is calculated 
by solving the local energy balance equation (\ref{steady energy balance}) (see details in Appendix \ref{app:steady solutions}). Specifically, the left panel, the middle panel and the right panel are for $r=5, 20, 40 R_{\rm S}$ respectively. 

From the upper panel of Figure \ref {Lum_Scurve}, it can be seen that the luminosity (blue and orange) decreases stably before 508 days after injecting matter at $R_{\rm out}$. The corresponding evolution of the accretion disk in $T_{\rm eff}-\Sigma$ diagram at $r=5, 20, 40 R_{\rm S}$ are shown in blue and orange in the lower left, middle and right panels of Figure \ref {Lum_Scurve} respectively. It can be seen that during this time the accretion disk is located in the upper branch (high state) of the S-shaped curve in $T_{\rm eff}-\Sigma$ diagram. After 508 days, corresponding to the
mass injection rate $\sim 1.92\dot M_{\rm Edd}$, the luminosity begins to oscillate until the end of the calculation of $\sim 3500$ days, i.e., the green, red, purple and pink lines in the upper panel of Figure \ref {Lum_Scurve}.
This oscillation can be clearly reflected in the limit cycle behaviors in $T_{\rm eff}-\Sigma$ diagram. One can see the lower panel of Figure \ref {Lum_Scurve} for the detailed evolution of the accretion disk at $r=5, 20, 40 R_{\rm S}$ respectively.

The oscillation is caused by the radiation pressure instability, the nature of which can be understood as follows.
When the accretion rate increases in a standard disk, the disk goes from being gas pressure ($p_{\rm{gas}}=\frac{k}{\mu m_{\rm{p}}}\rho T$) dominated to being radiation pressure ($p_{\rm{rad}}=\frac{a}{3}T^{4}$) dominated, which is viscously and thermally unstable. A small increase in temperature leads to a large increase in pressure, and hence a large increase in heating since the viscous stress $T_{\rm{r\varphi}}$ is assumed to be proportional to the total pressure. The radiative cooling is insufficient to keep up with the heating, so the disk is quickly heated up, and there is a runaway heating. However, as the accretion rate increases, the horizontal advection of the energy becomes important, which acts as a dominant cooling mechanism and can stabilize the disk. Thus, an upper stable branch with positive slope exists on the thermal equilibrium curve in $T_{\rm eff}-\Sigma$ diagram. The overheated disk jumps to the upper branch of the S-shaped curve instead of being destroyed. 
In our calculation, as we inject the fallback rate in the form of equation (\ref{M_dot_fb}) to the disk to simulate the TDE environment, the accretion rate in the disk begins on the upper stable branch of the S-shaped curve and gradually decreases to the upper turning point. Similarly, at this point, a small decease in temperature causes a large decrease in viscous heating. The disk quickly cools down and drops to the lower branch of the S-shaped curve. Then, since the mass supply rate at the outer disk is still at the unstable branch of the S-shaped curve, the limit cycle behavior occurs repeatedly and causes oscillations in the light curve. 

In the following, we test the effects of $M_{\rm BH}$, $\alpha$ and $R_{\rm out}$ on the light curves.

\begin{figure*}[ht!]
\centering
\includegraphics[scale=0.4]{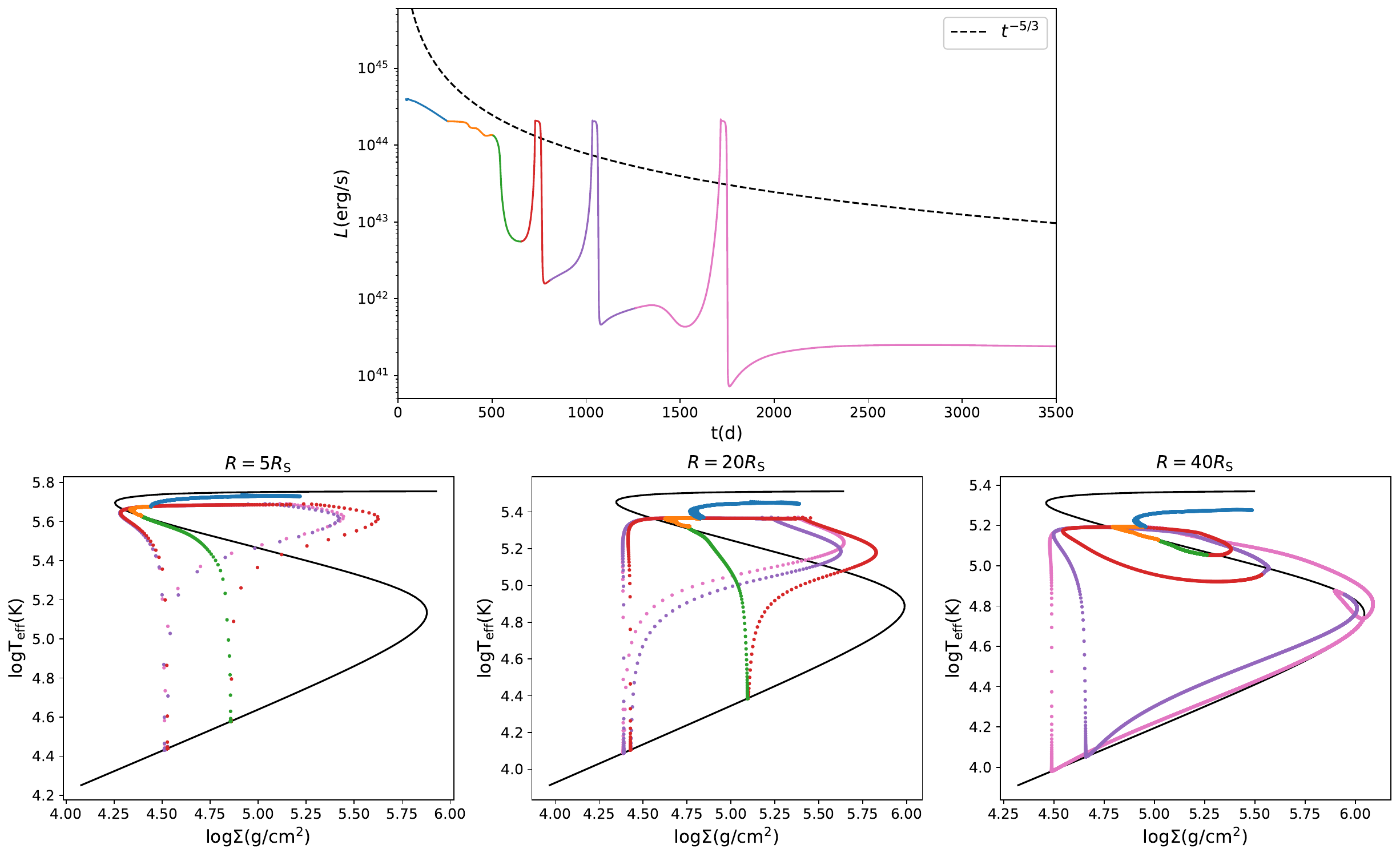}
\caption{Upper panel: The light curve. The colorful light curve refers to the results calculated with 
the time-dependent accretion disk by setting $M_{\rm{BH}}=10^{6}M_{\odot}$, $M_{\star} =M_{\odot}$, $R_{\star} =R_{\odot}$, $\alpha=0.01$, $\dot M_{\rm inject}=\dot M_{\rm fb}$ as equation (\ref{M_dot_fb}) and $R_{\rm out}=R_{\rm c}=47.5R_{\rm S}$ as equation (\ref{Rc}). The black dashed line is plotted by converting $\dot M_{\rm inject}$ to luminosity with $L=0.1\dot{M}_{\rm{inject}}c^{2}$. 
Lower panel: $T_{\rm eff}-\Sigma$ diagram. The black S-shaped curves in the left panel, middle panel
and the right panel are calculated with the thermal equilibrium equation (\ref{steady energy balance}) at $5R_{\rm{S}}$, $20R_{\rm{S}}$ and $40R_{\rm{S}}$ respectively. The evolution of accretion disk in $T_{\rm{eff}}-\Sigma$ diagram is plotted in multiple colors in the 
left panel, middle panel and the right panel for $5R_{\rm{S}}$, $20R_{\rm{S}}$ and $40R_{\rm{S}}$ respectively. The same color in the upper and lower panels represents the same evolution stage. }
\label{Lum_Scurve}
\end{figure*}

\subsection{Effect of $M_{\rm{BH}}$}
In Figure \ref{MBH}, we plot the light curves calculated with different $M_{\rm{BH}}$, i.e., $M_{\rm{BH}}=10^{6}M_{\odot}$, $10^{6.5}M_{\odot}$, $10^{7}M_{\odot}$ respectively. We set $\alpha=0.01$, $\dot M_{\rm inject}=\dot M_{\rm fb}$ as equation (\ref{M_dot_fb}) and $R_{\rm out}=R_{\rm c}$ as equation (\ref{Rc}). It can be seen that the outburst period becomes longer for larger $M_{\rm{BH}}$. This is because larger $M_{\rm{BH}}$ corresponds to longer viscous timescale, since the disk radius is larger for larger $M_{\rm{BH}}$, as shown in equation (\ref{Rc}). It can also be seen that for larger $M_{\rm{BH}}$, the steep drop in luminosity caused by the radiation pressure instability occurs at an earlier time. This is because $\dot{M}_{\rm{peak}}/\dot{M}_{\rm{Edd}}$ is smaller for larger $M_{\rm{BH}}$, as shown by (\ref{M_dot_peak}). As a result, the stable evolution stage before the instability occurs is shorter for larger $M_{\rm{BH}}$.

\begin{figure*}[ht!]
\centering
\includegraphics[scale=0.6]{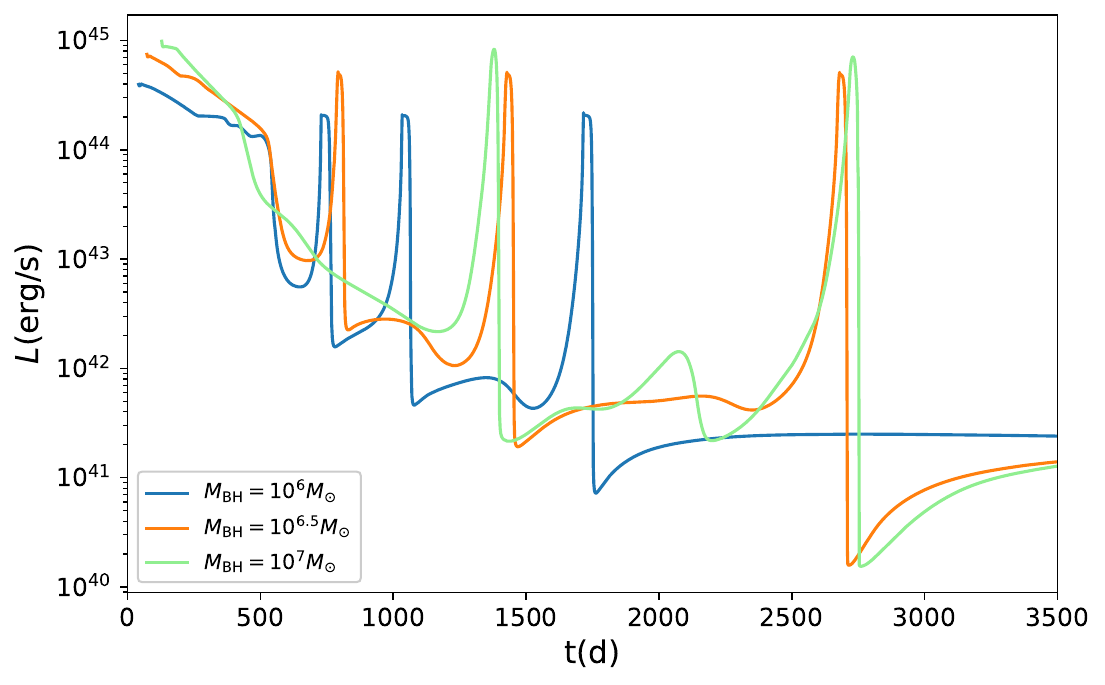}
\caption{Light curves for different $M_{\rm{BH}}$, i.e., $M_{\rm{BH}}=10^{6}M_{\odot}$, $10^{6.5}M_{\odot}$ and $10^{7}M_{\odot}$. We set $\alpha=0.01$, $\dot M_{\rm inject}=\dot M_{\rm fb}$ as equation (\ref{M_dot_fb}) and $R_{\rm out}=R_{\rm c}$ as equation (\ref{Rc}).}
\label{MBH}
\end{figure*}

\subsection{Effect of $\alpha$}
\label{effect of alpha}
In Figure \ref{alpha}, we plot the light curves calculated with different $\alpha$, i.e., $\alpha=0.001$, $0.01$, $0.1$ respectively. We set $M_{\rm{BH}}=10^{6}M_{\odot}$, $\dot M_{\rm inject}=\dot M_{\rm fb}$ as equation (\ref{M_dot_fb}) and $R_{\rm out}=R_{\rm c}$ as equation (\ref{Rc}).
It can be seen that $\alpha$ plays a crucial effect on the outburst periods. The outburst periods are shorter for larger $\alpha$. For $\alpha=0.1$ case, the periods of the outburst are as short as $12.5$ days in the early oscillation stage. This is because the mass transfer rate in the disk is faster for larger $\alpha$. The materials injected to the outer disk are quickly transferred inward, which accelerates the oscillations caused by the radiation pressure instability.

On the other hand, the oscillations 
are entirely suppressed for $\alpha=0.001$, as shown by the orange line in Figure \ref{alpha}. 
In order to better demonstrate the evolution of the disk and its light curve for this special case, in the upper panel of Figure \ref{Lum_Scurve_no_osci}, we plot the light curve for $\alpha=0.001$ in different colors. As a comparison, we  plot the mass injection rate (converted to luminosity by assuming $L=0.1\dot{M}_{\rm{inject}}c^{2}$)
as a function of time with black dashed line. Similar to that in Figure \ref{Lum_Scurve}, we plot the corresponding evolution of the disk in $T_{\rm{eff}}-\Sigma$ diagram at some fixed radius in the lower panel of Figure \ref{Lum_Scurve_no_osci}, together with the S-shaped curves plotted in black solid lines for comparison. Specifically, the left panel, the middle panel and the right panel are for $r=5$, $20$, $40R_{\rm{S}}$ respectively.
The same color in the upper and lower panels represents the same evolution stage. 

From the upper panel of Figure \ref{Lum_Scurve_no_osci}, it can be seen that before 540 days, first the luminosity decays stably as shown in blue, then the decay of luminosity becomes flat as shown in orange. From the lower panel of Figure \ref{Lum_Scurve_no_osci}, the corresponding evolution of the accretion disk in $T_{\rm{eff}}-\Sigma$ diagram shows that the flat decay in luminosity occurs when the disk evolves to near the upper turning point of the S-shaped curve, as shown in orange. At about 540 days, the radiation pressure instability occurs and the luminosity quickly decays as shown in green and red lines. The decay of this time is much steeper than the stable phase.
Finally the light curve settles to a roughly constant phase until the end of the calculation of $\sim 3500$ days, as shown in purple. From the $T_{\rm{eff}}-\Sigma$ diagram in the lower panel, it can be seen that at $r=5$ and $20R_{\rm{S}}$ (left and middle panel), $T_{\rm{eff}}$ nearly remains unchanged after dropping to the lower stable branch of S-shaped curves. In the right panel at $r=40R_{\rm{S}}$, $T_{\rm{eff}}$ only increases slightly (purple) after dropping to the lower branch. The little change in $T_{\rm{eff}}$ in the low state can be explained by the low mass transfer rate for such small $\alpha$. Few materials are transferred to the inner disk, thus $T_{\rm{eff}}$ at each disk radius remains almost constant. As a result, the total luminosity of the disk also remains constant during this time. Light curves with this shape can be responsible for some observations in TDEs as will be discussed in Section \ref{Application}.

\begin{figure*}[ht!]
\centering
\includegraphics[scale=0.6]{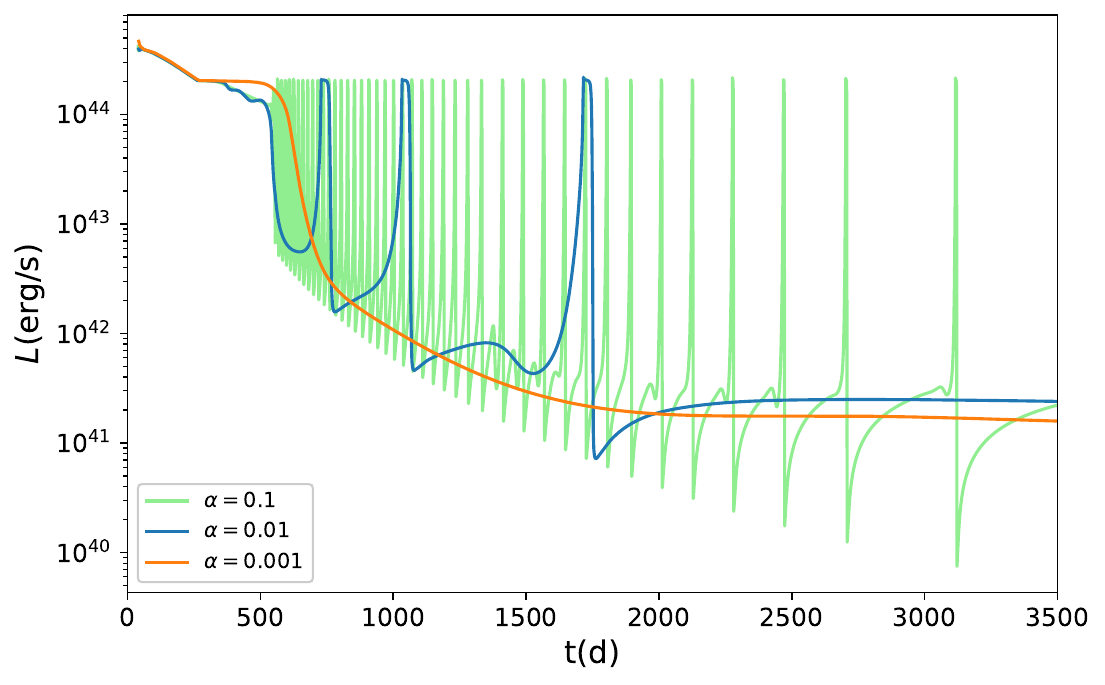}
\caption{ Light curves for different $\alpha$, i.e., $\alpha=0.001$, $0.01$ and $0.1$. We set $M_{\rm{BH}}=10^{6}M_{\odot}$, $\dot M_{\rm inject}=\dot M_{\rm fb}$ as equation (\ref{M_dot_fb}) and $R_{\rm out}=R_{\rm c}$ as equation (\ref{Rc}).}
\label{alpha}
\end{figure*}

\begin{figure*}[ht!]
\centering
\includegraphics[scale=0.4]{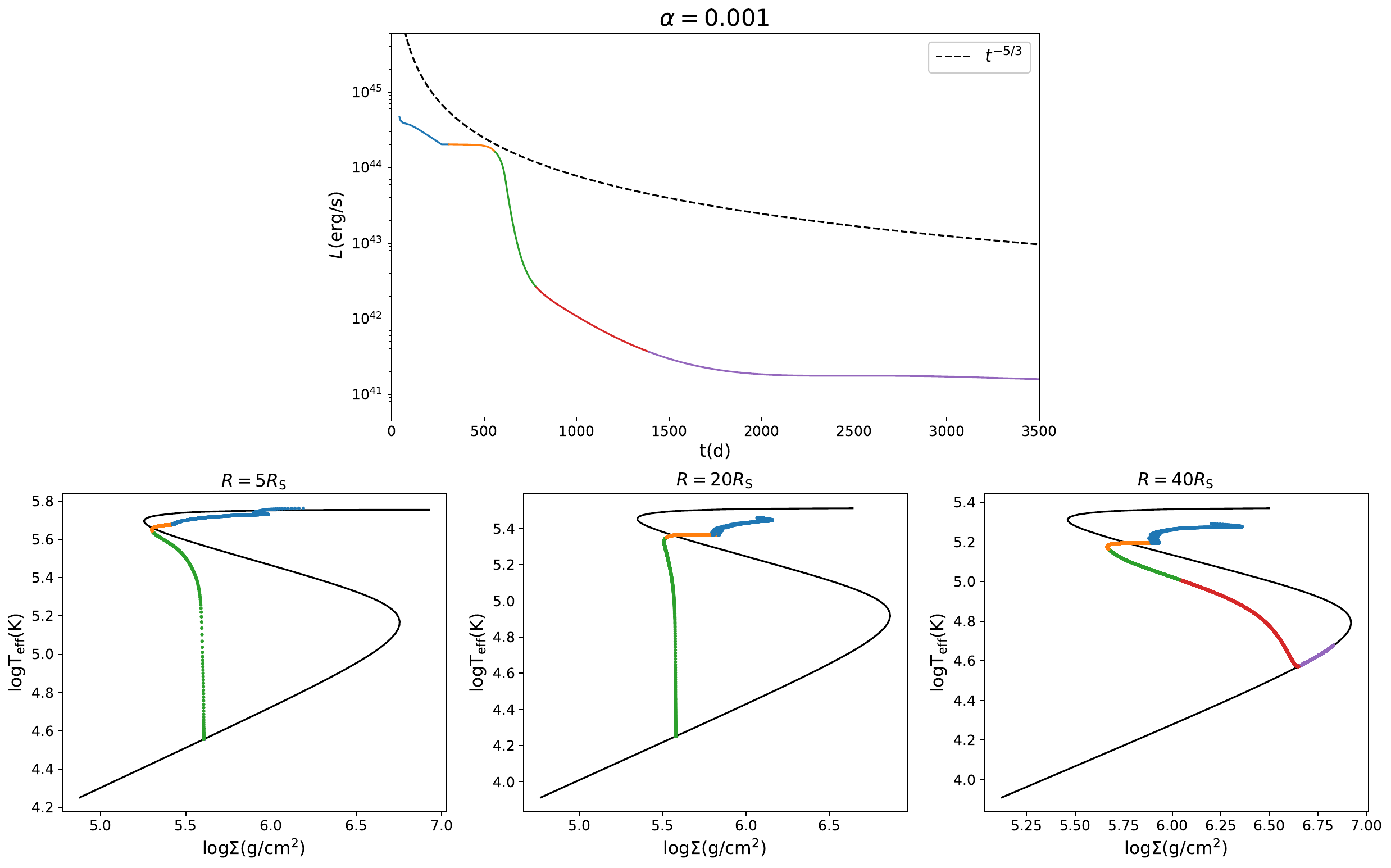}
\caption{
Upper panel: The light curve. The colorful light curve refers to the $\alpha=0.001$ case in Figure \ref{alpha}. The black dashed line is plotted by converting $\dot M_{\rm inject}$ to luminosity with $L=0.1\dot{M}_{\rm{inject}}c^{2}$. 
Lower panel: $T_{\rm eff}-\Sigma$ diagram. The black S-shaped curves in the left panel, middle panel
and the right panel are calculated with the thermal equilibrium equation (\ref{steady energy balance}) at $5R_{\rm{S}}$, $20R_{\rm{S}}$ and $40R_{\rm{S}}$ respectively. The evolution of accretion disk in $T_{\rm{eff}}-\Sigma$ diagram is plotted in multiple colors in the 
left panel, middle panel and the right panel for $5R_{\rm{S}}$, $20R_{\rm{S}}$ and $40R_{\rm{S}}$ respectively. The same color in the upper and lower panels represents the same evolution stage.
} 
\label{Lum_Scurve_no_osci}
\end{figure*}

\subsection{Effect of $R_{\rm{out}}$}
\label{effect of Rout}
In Figure \ref{Rout}, we plot the light curves calculated with different $R_{\rm{out}}$, i.e., $R_{\rm{out}}=0.5R_{\rm{c}}, R_{\rm{c}} $ and $2R_{\rm{c}}$ respectively. We set $M_{\rm{BH}}=10^{6}M_{\odot}$, $\dot M_{\rm inject}=\dot M_{\rm fb}$ as equation (\ref{M_dot_fb}) and $\alpha=0.01$. It can be seen that the steep drop in luminosity caused by the radiation pressure instability occurs earlier for larger $R_{\rm{out}}$. This can be clearly explained by Figure \ref{S_curve}, where we plot the S-shaped curves at different radii. The S-shaped curve shifts to the upper right in $T_{\rm{eff}}-\Sigma$ diagram for larger radius. For disk with larger $R_{\rm{out}}$, the declining $\dot{M}_{\rm{inject}}$ drops to the unstable branch of the S-shaped curve at an earlier time, thus the instability is triggered earlier. Moreover, the outburst periods are shorter for smaller $R_{\rm{out}}$.  This is because for disks with smaller $R_{\rm{out}}$, the injected mass in the outer disk takes less time to be transferred inward, leading to faster oscillations in the light curve once the instability occurs.

For $R_{\rm{out}}=2R_{\rm{c}}$, the outbursts are nearly completely suppressed. There's only a small bump at $\sim 1500$ days in the light curve, which becomes flat after 2000 days. Similar to the $\alpha=0.001$ case in Section \ref{effect of alpha}, we also plot the light curve for $R_{\rm{out}}=2R_{\rm{c}}$ in different colors in the upper panel of Figure \ref{Lum_Scurve_no_osci_Rout}, together with the evolution of the accretion disk in $T_{\rm{eff}}-\Sigma$ diagram in the lower panel. Specifically, the left, middle and right panels are for $r=10$, $50$, $80R_{\rm{S}}$ respectively. Other settings in Figure \ref{Lum_Scurve_no_osci_Rout} are the same as in Figure \ref{Lum_Scurve_no_osci}. From the lower panel of Figure \ref{Lum_Scurve_no_osci}, one can see a small limit cycle behavior near the lower turning point of the S-shaped curve at $r=80R_{\rm{S}}$ (right panel), which corresponds to the small bump in the light curve at $\sim 1500$ days. After dropping to the lower stable branch, $T_{\rm{eff}}$ at $10R_{\rm{S}}$ nearly remains constant, and $T_{\rm{eff}}$ at $50R_{\rm{S}}$ exhibits very small rise (purple). The slight change in $T_{\rm{eff}}$ in the inner disk is due to the longer time required for material to transfer inward for disk with larger $R_{\rm{out}}$. Consequently, the light curve shows small changes in the low state, especially after 2000 days until the end of the calculation.

\begin{figure*}[ht!]
\centering
\includegraphics[scale=0.6]{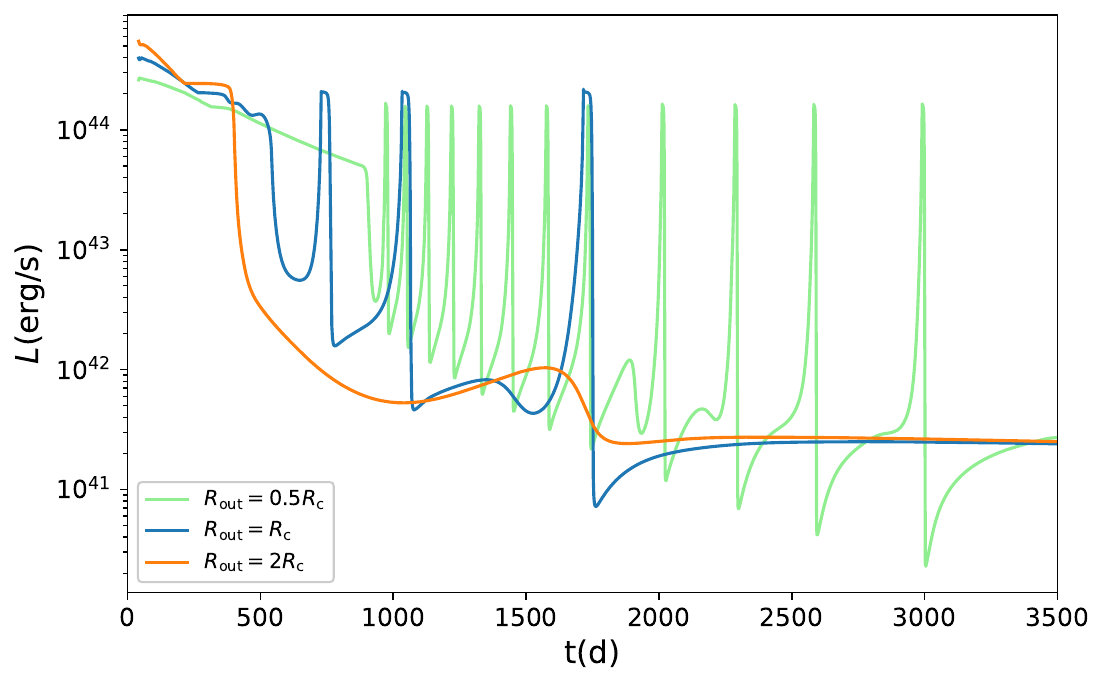}
\caption{ Light curves for different $R_{\rm{out}}$, i.e., $R_{\rm{out}}=0.5R_{\rm{c}}$, $R_{\rm{c}}$ and $2R_{\rm{c}}$, where $R_{\rm{c}}$ is calculated as equation (\ref{Rc}). We set $M_{\rm{BH}}=10^{6}M_{\odot}$, $\alpha=0.01$ and $\dot M_{\rm inject}=\dot M_{\rm fb}$ as equation (\ref{M_dot_fb}).}
\label{Rout}
\end{figure*}

\begin{figure*}[ht!]
\centering
\includegraphics[scale=0.4]{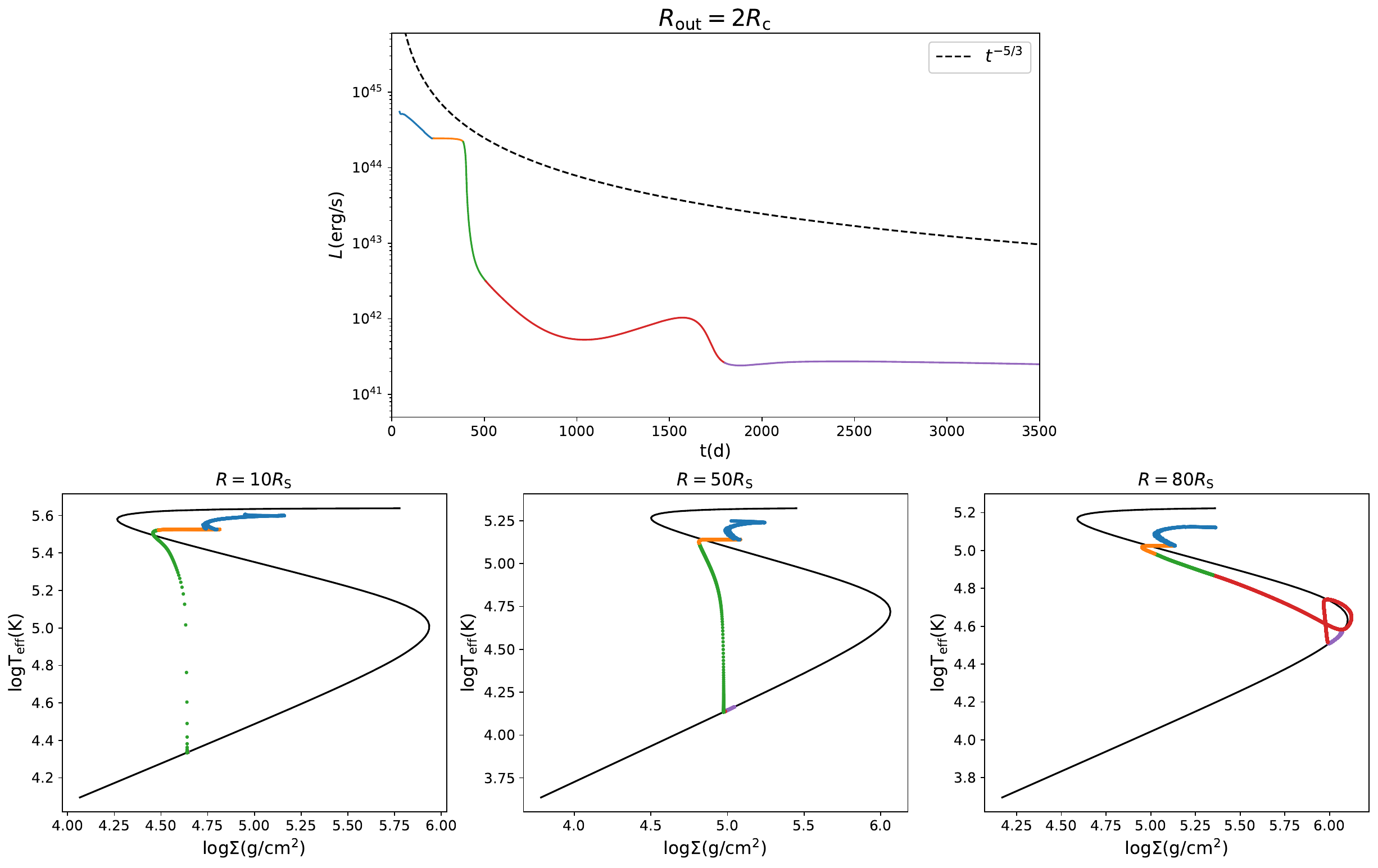}
\caption{
Upper panel: The light curve. The colorful light curve refers to the $R_{\rm{out}}=2R_{\rm{c}}$ case in Figure \ref{Rout}. The black dashed line is plotted by converting $\dot M_{\rm inject}$ to luminosity with $L=0.1\dot{M}_{\rm{inject}}c^{2}$. 
Lower panel: $T_{\rm eff}-\Sigma$ diagram. The black S-shaped curves in the left panel, middle panel
and the right panel are calculated with the thermal equilibrium equation (\ref{steady energy balance}) at $10R_{\rm{S}}$, $50R_{\rm{S}}$ and $80R_{\rm{S}}$ respectively. The evolution of accretion disk in $T_{\rm{eff}}-\Sigma$ diagram is plotted in multiple colors in the 
left panel, middle panel and the right panel for $10R_{\rm{S}}$, $50R_{\rm{S}}$ and $80R_{\rm{S}}$ respectively. The same color in the upper and lower panels represents the same evolution stage.
}
\label{Lum_Scurve_no_osci_Rout}
\end{figure*}

\section{Application to Observations}
\label{Application}
According to our calculations in Section \ref{Results_section}, the light curves of the time-dependent accretion disk in TDEs have various features under different parameter settings. Many of the light curves exhibit oscillations induced by radiation pressure instability, while there are special cases showing a complete absence of oscillations, as shown in the orange curve in Figure \ref{alpha} and Figure \ref{Rout}, which decline stably at the beginning of evolution, then drop due to radiation pressure instability and finally become flat in late times. With regard to the observations, many of the light curves do not show large-scale oscillations, especially in the UV/optical bands.
We note that several observed UV/optical light curves decay steeply in the early times and become flat in the late times (\cite{Velzen2019ApJ}), which are very similar to our calculated light curves with oscillations totally suppressed. In this work, we apply our model to explain the observed UV/optical light curves in these kinds of TDEs. The potential application of our model to some oscillations in light curves observed in TDEs will be discussed in Section \ref{Discuss_section}.

\subsection{Photosphere}
\label{Photosphere_section} 
Considering the UV/optical emissions in TDEs, the luminosity detected in the early times are always higher than that expected from an accretion disk.
Meanwhile, the estimated size of UV/optical emission region is roughly 1-2 orders of magnitude larger than the theoretical circularization radius $R_{\rm c}$. These emissions could be explained by the `reprocessing' model, with the X-ray/EUV radiation from the accretion disk being reprocessed into UV/optical band by a surrounding optically thick envelope or outflows (\cite{Loeb_Ulmer1997ApJ}, \cite{Dai2018ApJ}, \cite{Qiao2025MNRAS.539.3473Q}). 
Based on this point of view, in this work we construct a photosphere model to convert the disk luminosity into single-band UV/optical emissions, following a similar scheme proposed by \cite{Mockler2019ApJ}.
 
Assuming that the photosphere radiates as a blackbody, 
the observed flux can be written as,
\begin{equation}
    F_{\nu}=\frac{2\pi h\rm\nu^{3}}{c^{2}} 
 \frac{1}{\rm{exp}(h\nu/kT_{\rm{eff,ph}})-1} \frac{R_{\rm{ph}}^{2}}{D^{2}},  \label{Fnu_ph}
\end{equation}
where $D$ is the distance from the source, $h$ is the Planck constant,  $T_{\rm{eff,ph}}$ is the effective temperature of the photosphere, and $R_{\rm{ph}}$ is the photosphere radius.
The effective temperature of the photosphere can be expressed as,
\begin{equation}
    T_{\rm{eff,ph}}=\left(\frac{L}{4\pi \sigma R_{\rm{ph}}^{2}} \right )^{1/4},  \label{Teff_ph}
\end{equation}
where $L$ is the total luminosity of the disk. We have followed the assumption that all of the radiation from the disk is efficiently thermalized at the scale of the photosphere radius. 
We model the photosphere radius $R_{\rm{ph}}$ as a power-law function of the luminosity $L$. Given that the dependence of $R_{\rm{ph}}$ on $L$ in early times and late times may be different, we adopt a piecewise formula, which can be written as
\begin{equation}
    R_{\rm{ph}}=
\begin{cases}
R_{\rm{ph0}}\left( L/L_{\rm{Edd}} \right)^{l},t\le t_{\rm{c}}\\
R_{\rm{ph}c}\left( L/L_{\rm{c}} \right)^{l_{\rm{c}}}, \quad t>t_{\rm{c}},
\end{cases} \label{Rph}
\end{equation}
where $R_{\rm{ph0}}$, $l$,  $t_{\rm{c}}$ and $l_{\rm{c}}$ are four parameters to be determined. $R_{\rm{phc}}=R_{\rm{ph}}(t_{\rm{c}})$ and $L_{\rm{c}}=L(t_{\rm{c}})$ are the photosphere radius and luminosity at the turning time $t_{\rm{c}}$, respectively, which can be calculated with the upper expression in equation (\ref{Rph}) once $R_{\rm{ph0}}$, $l$ and $t_{\rm{c}}$ are specified. We define the turning point $t_{\rm{c}}$ as the time when $L/L_{\rm{Edd}}$ evolves to a certain value $L_{\rm{c}}/L_{\rm{Edd}}$. As for the other three parameters $R_{\rm{ph0}}$, $l$ and $l_{\rm{c}}$, we constrain them by fitting the multi-band UV detections with the flux calculated by equation (\ref{Fnu_ph}) combining equation (\ref{Teff_ph}) and (\ref{Rph}).

\subsection{Fitting results}
\subsubsection{Fitting procedure}
To fit the UV/optical observations, first we need to calculate the light curve from the time-dependent accretion disk, then we convert the disk luminosity into UV/optical flux with the photosphere model described in Section \ref{Photosphere_section}, which is used to fit the UV/optical data. In calculating the disk light curves, we rewrite equation (\ref{M_dot_fb}) as $\dot{M}_{\rm{fb}}=\dot{M}_{\rm{ini}}\left (1+t/t_{\rm{fb}} \right )^{-5/3}$, such that the luminosity peak is at $t=0$ in our calculations. We take $\dot{M}_{\rm{ini}}$ as a parameter since it's not always equal to $\dot{M}_{\rm{peak}}$, as shown in equation (\ref{M_dot_peak}). This is because some of the fallback materials could be blown off by shocks generated in the stream-stream collisions during circularization (\cite{Jiang2016ApJ}), a process that involves much uncertainty. We set $R_{\rm{out}}=R_{\rm{c}}$ as equation (\ref{Rc}), to consider a standard case. In this setup, the parameters we need to determine for the time-dependent disk are $M_{\rm{BH}}$, $\alpha$ and $\dot{M}_{\rm{ini}}$. 
In this work, these three parameters cannot be constrained by fitting the observed UV/optical light curves. This is because the detailed fit requires an exploration of the full parameter space of our time-dependent accretion disk model, which entails substantial time and computational resources, and is not currently available. Instead, we first calculate several disk light curves with different sets of $M_{\rm{BH}}$, $\alpha$, $\dot{M}_{\rm{ini}}$, and fit them to the observed UV/optical emissions with the photosphere model. Then, by comparing the fitting results, we can determine the best set of these parameters for the time-dependent disk.

After calculating the disk light curve with a given set of $M_{\rm{BH}}, \alpha$ and $\dot{M}_{\rm{ini}}$, we fit the UV/optical data by equation (\ref{Fnu_ph}) combining equation (\ref{Teff_ph}) and (\ref{Rph}). Given that some sources lack detection of the early rising phase and the corresponding peak in the optical light curves, the detection time of the first UV/optical data point could be later than the optical peak time, i.e., $t=0$ in our model. Here, we define $t_{\rm{o}}$ as the time when the UV/optical flux is first detected with respect to $t=0$. It turns out to be difficult to incorporate $t_{\rm{o}}$ and the other time parameter $t_{\rm{c}}$ into fitting, so we have to treat them separately from other parameters in equation (\ref{Rph}). Since we have defined $t_{\rm{c}}$ to be the time when $L/L_{\rm{Edd}}$ evolves to a certain value $L_{\rm{c}}/L_{\rm{Edd}}$, we can determine $t_{\rm{c}}$ by specifying $L_{\rm{c}}/L_{\rm{Edd}}$. 
When $t_{\rm{c}}$ is given, the $t\le t_{\rm{c}}$ and $t> t_{\rm{c}}$ parts can be fitted individually. For the $t\le t_{\rm{c}}$ part, we adjust the value of $t_{\rm{o}}$ to achieve the best fit, from which $R_{\rm{ph0}}$ and $l$ are determined.  Then, we fit the late-time detections ($t> t_{\rm{c}}$) to get the value of $l_{\rm{c}}$. The fitting procedure of the photosphere model is conducted with Markov Chain Monte Carlo (MCMC) method(\cite{Foreman-Mackey2013PASP}).

\subsubsection{Fitting to ASASSN-15oi and ASASSN-14ae}
We choose ASASSN-15oi(\cite{Holoien2016MN_ASASSN-15oi_1}, \cite{Gezari2017ApJ_ASASSN-15oi_2}) and ASASSN-14ae(\cite{Holoien2014MN_ASASSN-14ae_1}, \cite{Brown2016MN_ASASSN-14ae_2}) as two examples for the application of our model. Both sources are bright in UV and optical bands. 
We mainly fit the UV detections which are significantly above the host galaxy contributions. 
In the upper panels of Figure \ref{15oi_lc_Rph}, we plot the best-fit UV light curves of ASASSN-15oi. Both the calculated flux and the data are converted to AB magnitude. The light curves are calculated with our disk model by setting  $M_{\rm{BH}}=5\times10^{6}M_{\odot}, \alpha=0.008$ and  $\dot{M}_{\rm{ini}}=2\dot{M}_{\rm{Edd}}$, then fitted to the UV data with the photosphere model. The value of $M_{\rm{BH}}/M_{\odot}$ we determined here is within 0.5 dex of $10^{7.1}$ estimated from BH-mass-to-bulge-mass ($M_{\rm{BH}}-M_{\rm{bulge}}$) relation (\cite{Holoien2016MN_ASASSN-15oi_1}). This is reasonable since the intrinsic scatter in $M_{\rm{BH}}-M_{\rm{bulge}}$ relation is $\sim 0.5$ dex (\cite{McConnell_Ma2013ApJ}, \cite{Mockler2019}).
The UV data (\cite{Holoien2016MN_ASASSN-15oi_1},\cite{Gezari2017ApJ_ASASSN-15oi_2}) is host-subtracted, and is plotted with $\rm{MJD}-57248.2+t_{\rm{o}}$, where $\rm{MJD}$ $57248.2$ is the epoch of discovery, and the detection time $t_{\rm{o}}$ is set at $28.4$ days. The left and right panels in Figure \ref{15oi_lc_Rph} refer to the fitting results by setting $L_{\rm{c}}/L_{\rm{Edd}}=0.003$ and $L_{\rm{c}}/L_{\rm{Edd}}=0.001$, i.e., $t_{\rm{c}}=265\rm{d}$ and $t_{\rm{c}}=403\rm{d}$ in equation (\ref{Rph}) respectively. It can be seen that the long-term evolution of the UV light curves spanning $\sim 650$ days can be well explained by our disk model together with a photosphere. The setting of $L_{\rm{c}}/L_{\rm{Edd}}$ slightly affects the shape of the late-time UV light curves, and does not affect the fitting performance for this source.

In the lower panels of Figure \ref{15oi_lc_Rph}, we show the evolution of $R_{\rm{ph}}$ of ASASSN-15oi. The pink solid lines are the results calculated with equation (\ref{Rph}), the parameters of which are determined by fitting the UV light curve. As for comparison, we plot the disk outer radius $R_{\rm{out}}=R_{\rm{c}}=16R_{\rm{S}}$ in green dashed line. It can be seen that $R_{\rm{ph}}$ is much larger than $R_{\rm{out}}$ in the early times, indicating significant reprocessing. In the late times, $R_{\rm{ph}}$ recedes to nearly the same size as the disk, indicating that the UV flux of the photosphere model is similar to that of the disk model. We show the comparison between the UV light curves of the photosphere (solid lines) and that of the disk (dashed lines) in Figure \ref{15oi_disk}. It can be seen that in the late times, the flux from the photosphere is consistent with the pure disk emissions. This is in agreement with several current works which show that the late-time optical/UV emissions in TDEs can be well modeled by an accretion disk (e.g., \cite{Mummery_Balbus2020MNRAS}, \cite{Alush_2025_2}).

In the upper panel of Figure \ref{14ae_lc_Rph}, we plot the best-fit UV light curves of ASASSN-14ae. Both the calculated flux and the data are converted to AB magnitude. The light curves are first calculated with the disk model by setting $M_{\rm{BH}}=3.5\times10^{6}M_{\odot}, \alpha=0.007$ and $\dot{M}_{\rm{ini}}=2\dot{M}_{\rm{Edd}}$, then fitted to the UV data with the photosphere model.
The value of $M_{\rm{BH}}/M_{\odot}$ determined here is within 0.5 dex of $10^{6.8}$ estimated from $M_{\rm{BH}}-M_{\rm{bulge}}$ relation (\cite{Holoien2014MN_ASASSN-14ae_1}). 
The UV data (\cite{Holoien2014MN_ASASSN-14ae_1}, \cite{Brown2016MN_ASASSN-14ae_2}) is host-subtracted and is plotted with $\rm{MJD}-56682.5+t_{\rm{o}}$, where $\rm{MJD}$ $56682.5$ is the epoch of discovery, and the detection time $t_{\rm{o}}$ is set at $39$ days with respect to $t=0$. The left and right panels in Figure \ref{14ae_lc_Rph} refer to the fitting results by setting $L_{\rm{c}}/L_{\rm{Edd}}=0.003$ and $L_{\rm{c}}/L_{\rm{Edd}}=0.001$, i.e., $t_{\rm{c}}=195\rm{d}$ and $t_{\rm{c}}=288\rm{d}$ in equation (\ref{Rph}) respectively. It can be seen that the long-term evolution of the observed UV light curves spanning $\sim 750$ days can be reproduced by our disk model together with a photosphere. The setting of $L_{\rm{c}}/L_{\rm{Edd}}$ slightly affects the shape of the late-time UV light curves, and does not affect the fitting performance for this source. 

 In the lower panel of Figure \ref{14ae_lc_Rph}, we show the evolution of $R_{\rm{ph}}$ of ASASSN-14ae in pink lines. The green dashed lines show the disk outer radius $R_{\rm{out}}=R_{\rm{c}}=20R_{\rm{S}}$ as for comparison. 
It can be seen that $R_{\rm{ph}}$ is much larger than $R_{\rm{out}}$ in the early times, indicating strong reprocessing. In the late times, $R_{\rm{ph}}$ recedes to $\sim 3R_{\rm{out}}$. We show the comparison between the UV light curves of the photosphere (solid lines) and that of the disk (dashed lines) in Figure \ref{14ae_disk}. In the late times, while the UV flux of the photosphere model and the disk model is similar, the photosphere model appears to better fit the data. 
This suggests that the reprocessing could still play a role in the late times, 
which might require a mechanism such as magnetically-driven wind from a thin disk to continuously replenish the photosphere. On the other hand, in this work we have fixed $R_{\rm{out}}=R_{\rm{c}}$. Further consideration of a spreading disk will naturally lead to  
larger emitting area for optical/UV photons. In this way, the UV flux from the disk could better match the data, as demonstrated by several current works that successfully model the late-time optical/UV plateau with a spreading disk (e.g., \cite{Mummery_Balbus2020MNRAS}, \cite{Alush_2025_2}). 

The UV light curves of ASASSN-15oi and ASASSN-14ae show similar features. They decline steeply in the early times and become flat in the late times.  Although the photosphere plays a role in modulating the magnitude of the UV flux, it does not significantly affect the overall shape of the light curve, since the evolution of $R_{\rm{ph}}$ follows that of the luminosity, as shown in equation (\ref{Rph}). So we can understand the evolution of the UV flux according to our analysis of the disk luminosity in Section \ref{effect of alpha}. The decline in flux in the early $\sim 300$ days for these two sources is caused by the radiation pressure instability, during which the accretion rate quickly decreases and the disk transforms from a slim disk to a thin disk. After that, the disk settles on the lower stable branch of the S-shaped curve, with $T_{\rm{eff}}$ on each disk radius nearly remaining unchanged, as shown in the lower panels of Figure \ref{Lum_Scurve_no_osci}. This is because the mass transfer rate in the disk is low due to the small $\alpha$ and low $\dot{M}_{\rm{inject}}$ at this time. Thus, the UV flux is almost constant in the late times, which naturally explains the flattened feature of these two sources. This can be impressive since the luminosity of the disk remains constant in the late times but does not follow the decline of the $t^{-5/3}$ fallback rate. While a flat disk light curve can be obtained by calculating a viscously spreading thin disk without considering the fallback process, which is currently applied to interpret the late-time plateau(\cite{Velzen2019ApJ}, \cite{Mummery_Balbus2020MNRAS}), we show that this feature can be reproduced when accounting for the full evolution of the disk with continuous mass injection in the form of the fallback rate.

\begin{figure*}[ht!]
\centering
\includegraphics[scale=0.5]{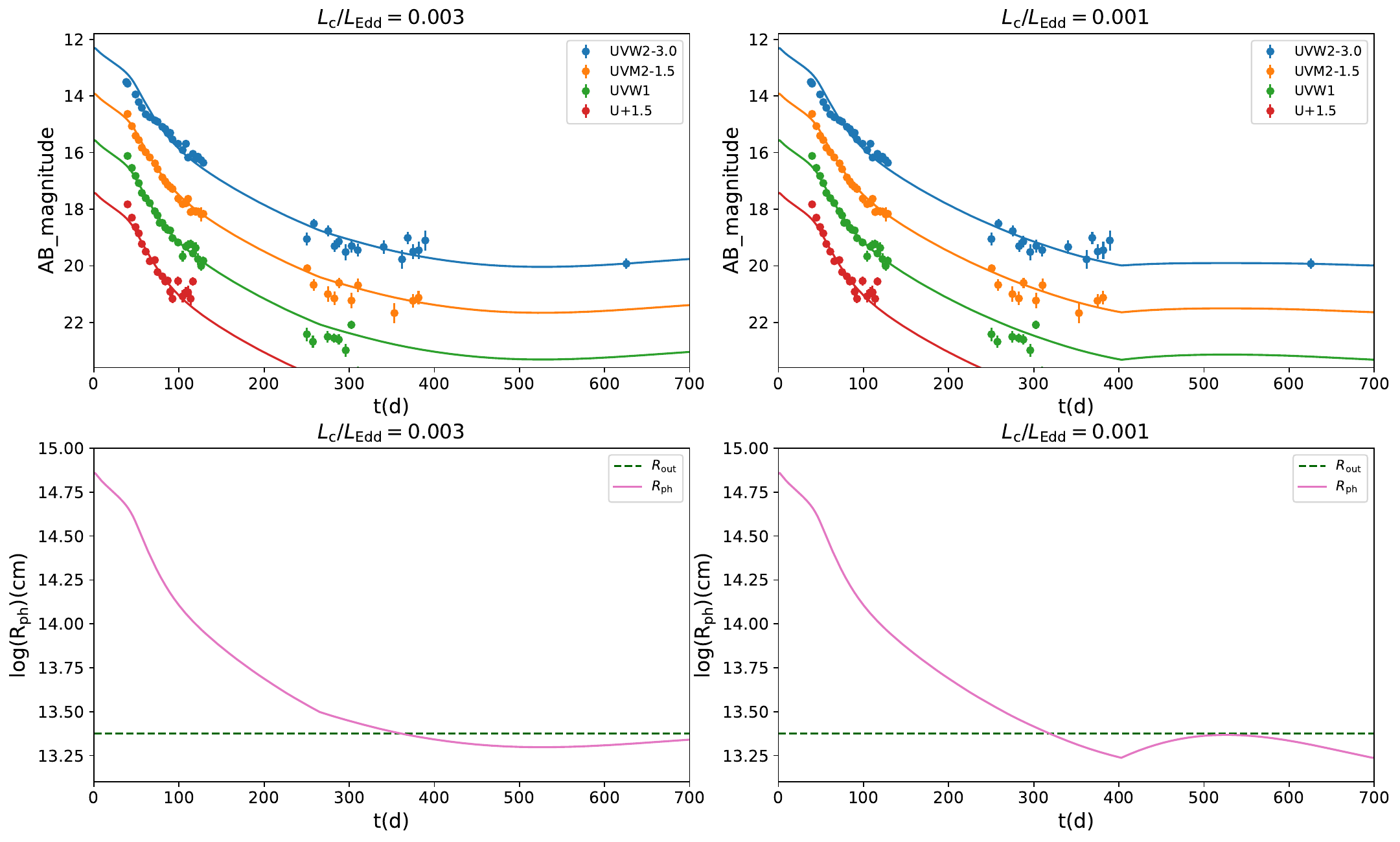}
\caption{Upper panels: The best-fit UV light curves of ASASSN-15oi. The solid light curves are the results calculated with our disk model by setting  $M_{\rm{BH}}=5\times10^{6}M_{\odot}, \alpha=0.008$, $\dot{M}_{\rm{ini}}=2\dot{M}_{\rm{Edd}}$, and fitted to the observation data with the photosphere model described in Section \ref{Photosphere_section}. The UV data (\cite{Holoien2016MN_ASASSN-15oi_1},\cite{Gezari2017ApJ_ASASSN-15oi_2}) are plotted with $\rm{MJD}-57248.2+t_{\rm{o}}$, where $\rm{MJD}$ $57248.2$ is the epoch of discovery and the detection time $t_{\rm{o}}$ is set at $28.4$ days. The left panel and right panel show the light curves fitted by setting $L_{\rm{c}}/L_{\rm{Edd}}=0.003$ and $L_{\rm{c}}/L_{\rm{Edd}}=0.001$, i.e., $t_{\rm{c}}=265\rm{d}$ and $t_{\rm{c}}=403\rm{d}$, respectively.
Lower panels: The evolution of photosphere radius of ASASSN-15oi. The pink solid lines are the results calculated with equation (\ref{Rph}), the parameters of which are determined by fitting the UV light curves. The green dashed lines show the disk outer radius in calculating the disk light curve, where $R_{\rm{out}}=R_{\rm{c}}=16R_{\rm{S}}$. }
\label{15oi_lc_Rph}
\end{figure*}

\begin{figure*}[ht!]
\centering
\includegraphics[scale=0.5]{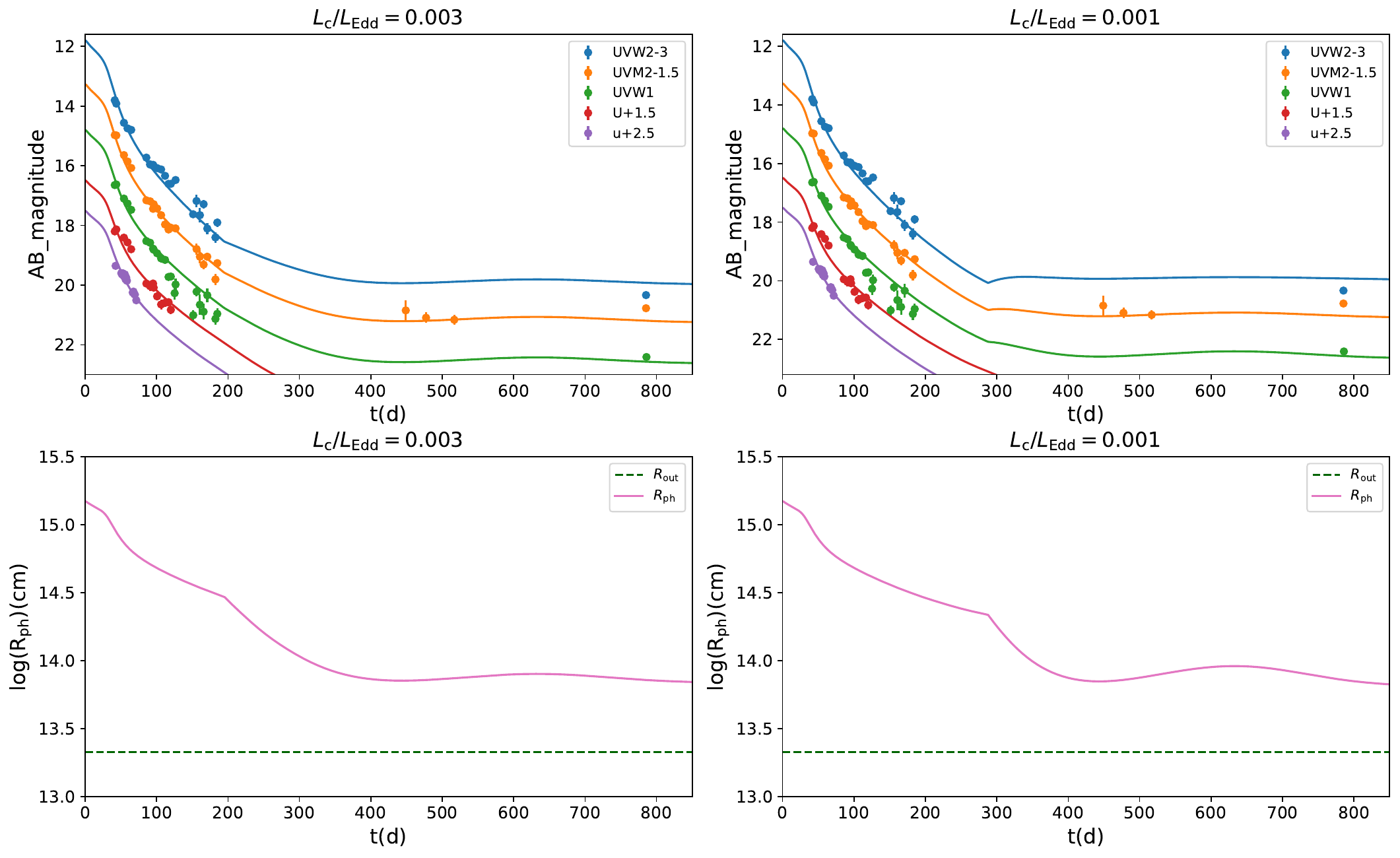}
\caption{Upper panel: The best-fit UV light curves of ASASSN-14ae. The solid light curves are the results calculated with our disk model by setting  $M_{\rm{BH}}=3.5\times10^{6}M_{\odot}, \alpha=0.007$, $\dot{M}_{\rm{ini}}=2\dot{M}_{\rm{Edd}}$, and fitted to the observation data with the photosphere model described in Section \ref{Photosphere_section}. The UV data (\cite{Holoien2014MN_ASASSN-14ae_1}, \cite{Brown2016MN_ASASSN-14ae_2}) are plotted with $\rm{MJD}-56682.5+t_{\rm{o}}$, where $\rm{MJD}$ $56682.5$ is the epoch of discovery and the detection time $t_{\rm{o}}$ is set at $39$ days. The left panel and right panel show the light curves fitted by setting $L_{\rm{c}}/L_{\rm{Edd}}=0.003$ and $L_{\rm{c}}/L_{\rm{Edd}}=0.001$, i.e., $t_{\rm{c}}=194.7\rm{d}$ and
$t_{\rm{c}}=287.6\rm{d}$, respectively. 
Lower panels: The evolution of photosphere radius of ASASSN-14ae. The pink solid lines are the results calculated with equation (\ref{Rph}), the parameters of which are determined by fitting the UV light curves. The green dashed lines show the disk outer radius in calculating the disk light curve, where $R_{\rm{out}}=R_{\rm{c}}=20R_{\rm{S}}$. }
\label{14ae_lc_Rph}
\end{figure*}

\section{Discussion}
\label{Discuss_section}
\subsection{Comparison with other works}
\label{Discuss_Comparison}
In this work, as stated in Section \ref{Disk model}, the outer boundary of the disk is chosen to have a stable solution, calculated by the thermal equilibrium equation (\ref{steady energy balance}) with $\dot{M}(t) =\dot{M}_{\rm{fb}}(t)$ at each time separately. This treatment is adapted from the studies of radiation pressure instability in X-ray binaries and active galactic nuclei (AGN) (\cite{Janiuk2002ApJ}, \cite{Sniegowska2022}, \cite{Sniegowska2023A&A}).
However, in calculating the disk in TDEs, this method has a disadvantage that, it prevents the self-consistent calculation of the viscous spreading of the outer disk. Consequently, we have to fix the radius of the outer boundary in our calculations. This will influence the calculation of realistic optical/UV emissions from the accretion disk, especially in the late times, when the optical/UV emissions in TDEs are generally thought to come from the disk (e.g., \cite{Mummery_Balbus2020MNRAS}, \cite{Guolo2024ApJ...966..160G_Ref1}, \cite{Alush_2025ApJ_1}). 
Nevertheless, in interpreting the optical/UV light curves, our photosphere model can equivalently account for some key observational consequences as that of the spreading disk, i.e., cooling and extending the emission region for producing the optical/UV emission. 
Thus, our approach remains valid for describing the long-term behavior of the optical/UV light curves, especially the early-time optical/UV emissions. For a more self-consistent treatment of the disk evolution, we should further investigate the disk spreading scenario with other factors considered, such as outflows and different descriptions of viscosity in the future work. 

In this work, we calculate the evolution of a 1D time-dependent disk with continuous mass injection at the outer boundary in the form of $\dot{M}_{\rm{inject}}\propto t^{-5/3}$. We include the advection term in our time-dependent energy conservation equation, thus we are able to simulate both the early-time slim disk phase and the late-time thin disk phase.
We obtain light curves that show radiation pressure instability in our calculations, and demonstrate how the light curves are affected by $M_{\rm{BH}}$, $\alpha$ and $R_{\rm{out}}$. These results are similar to the findings in \cite{Piro_2025} to some extent. However, in this paper, we use a 1D disk model to calculate the time-dependent disk, while \cite{Piro_2025} adopts a one-zone model. The advantage of a 1D model is that it provides more detailed description of the radial structure and evolution of the accretion disk.
 
Regarding the application to the observed UV light curves in TDEs, 
\cite{Piro_2025} compares the disk light curve that shows oscillations induced by the radiation pressure instability with the observed late-time data. However, the data are combined from different events and cannot be viewed as a real light curve.
In our work, we collect the UV light curves of two TDEs, ASASSN-15oi and ASASSN-14ae, which exhibit a steep decay in the early times and become flat in the late times. To model this long-term evolution, we adopt theoretical disk light curves in which oscillations are fully suppressed. These light curves are then converted into UV fluxes via our photosphere model and fitted to the observational data. 
We show that the long-term evolution of the observed light curves can be self-consistently explained by our model. In particular, we demonstrate that the steep decay in the UV/optical light curves of these two sources could be a result of the radiation pressure instability.

One deficiency of our model is that the late-time luminosity is too low to account for other TDEs with brighter late-time plateaus. Previous works modeling the late-time optical/UV light curves have suppressed the radiation pressure instability by invoking either different viscosity descriptions (\cite{Velzen2019ApJ}, \cite{Mummery_Balbus2020MNRAS}) or magnetic pressure(\cite{Alush_2025ApJ_1}, \cite{Kaur_2023MNRAS.524.1269K_Ref7}). These modifications prevent a steep drop in the light curves induced by the radiation pressure instability and thereby maintain a disk luminosity sufficient to explain the late-time UV/optical detections. For future improvements of our model, we will incorporate these factors to mitigate the drop in luminosity caused by the radiation pressure instability. Additionally, we will consider other proposed mechanisms, including fallback heating and the iron bump opacity as suggested by \cite{Piro_2025}. 

In previous works, the early-time decay and late-time flattening in the UV light curves in TDEs are treated separately, since they are generally expected to be dominated by different processes. The early-time decay is usually fitted by a photosphere model(\cite{Mockler2019}), and the late-time flattening is modeled by a spreading disk (\cite{Mummery2024MNRAS}, \cite{Alush_2025ApJ_1}). In this work, for the first time, we interpret the long-term UV light curves including both stages with a continuous evolution of the disk, together with a photosphere model. 
In our framework, the early steep decay is driven by the radiation pressure instability. Then the disk can remain in the lower branch of the S-shaped curve without continuing the limit-cycle behavior when $\alpha$ is small, producing a flat feature in the late-time disk light curves. Such disk behaviors greatly influence the evolution of the photosphere. In the early times, $R_{\rm{ph}}$ is much larger than $R_{\rm{out}}$, which could indicate strong reprocessing. As the disk luminosity decays, the photosphere recedes. In the late times, $R_{\rm{ph}}$ becomes comparable to $R_{\rm{out}}$, implying that the UV flux of the photosphere model is similar to that of the disk. This result is consistent with current understanding that the late-time UV emissions are mainly dominated by the disk.

\subsection{Improvements of the model}
\label{Disscuss_Improvements of the model}
In this work, as we calculate the evolution of the time-dependent disk in the standard TDE environment, $\dot{M}_{\rm{inject}}$ is initially super-Eddington, as shown by equation (\ref{M_dot_peak}). In a super-Eddington accretion disk with strong radiation pressure, a significant fraction of the matter is expected to be blown off forming outflow (\cite{Jiang2014ApJ.796.106J}, \cite{Dai2018ApJ}, \cite{Qiao2025MNRAS.539.3473Q}). Outflows could play a role in modulating the disk light curves. For example, it may shorten the periods and reduce the amplitude of the oscillations caused by radiation pressure instability(\cite{Janiuk_2015_IGR}). Moreover, outflows from the disk could naturally generate the optical thick reprocessing layer we used in interpreting the UV observations in Section \ref{Application}. However, in our 1D disk model, it's unable to predict the launch of wind since the disk is assumed to remain in hydrostatic equilibrium in the vertical direction (see equation (\ref{hydrostatic equilibrium})). To account for the effects of winds in our work, we should artificially introduce outflows into our disk model. For example, we can estimate the fraction of mass taken away by the winds according to our simulation work \cite{Qiao2025MNRAS.539.3473Q}. Then, we can add it to the equations describing the time-dependent disk in Section \ref{Disk model} to simulate the loss of mass, angular momentum and energy in the wind in our disk model. 

Besides outflows, another important factor we should take into consideration is the different forms of the fallback rate in TDEs. In this work, we just test the most standard case, i.e., setting $\dot{M}_{\rm{inject}}\propto t^{-5/3}$, where we have assumed that the star is fully disrupted and $\beta=1$. According to simulations of the fallback process, when a star is not fully disrupted in a TDE, the decay rate of $\dot{M}_{\rm{fb}}$ can be steeper than $t^{-5/3}$ (\cite{Guillochon2013ApJ}). Furthermore, the fallback rate is closely related to the polytropic model of the star and the stage of disruption. For a more precise treatment, we will adopt the fallback rate presented in the simulation work by \cite{Guillochon2013ApJ} and inject it into the disk, and consider the disruption of different types of stars. In this way, we are able to investigate various behaviors of the light curve in different TDE environments. 

In this paper, we mainly set $R_{\rm{out}}=R_{\rm{c}}$ as equation (\ref{Rc}) to consider the standard case. According to the results calculated with the one-zone disk model in \cite{Shen2014ApJ}, the disk may not be confined to the circularization radius but will spread out due to angular momentum conservation. As shown in section \ref{effect of Rout}, $R_{\rm{out}}$ larger than $R_{\rm{c}}$ could reduce the mass transfer rate in the disk and suppress the oscillations induced by radiation pressure instability. Thus, a spreading disk may account for the lack of oscillations in current observations in TDEs. In Section \ref{effect of Rout}, we only take $R_{\rm{out}}$ as a parameter and test its influence on the light curves. To further consider a spreading disk, we shall calculate the evolution of $R_{\rm{out}}$ instead of taking it as a fixed parameter, which will be studied in future work.  

\subsection{Applications to X-ray observations in the future}
In this work, we have mainly applied our disk model together with a photosphere to explaining the UV light curves in TDEs. In addition to the UV/optical emissions, our disk model has potential applications in interpreting the X-ray observations, since it is generally believed that the soft X-ray emissions in TDEs are produced by the accretion disk. For some X-ray selected TDEs that are faint in UV/optical band, we can explain their soft X-ray emissions simply with the flux calculated from the accretion disk. However, for some TDEs that are bright in both optical and X-ray bands, calculation of the X-ray flux could be complicated because the optical and X-ray emissions could be coupled with each other according to the reprocessing model. Some of the X-ray emissions are reprocessed into UV/optical band, thus the X-ray flux is greatly affected by the properties of the photosphere. 
The photosphere model we adopted in Section \ref{Photosphere_section} is incapable of producing the X-ray emissions, since we have made the assumption that the reprocessing layer covers the whole disk. 
According to simulations, the inner part of the disk (within $\sim 5R_{\rm{S}}$) is not enveloped by the optically thick reprocessing layer (\cite{Jiang2014ApJ.796.106J}). Most of the X-ray emissions from this inner part of the disk can escape and be radiated.
Therefore, we can modify the photosphere model to expose the inner disk to account for the UV/optical and X-ray emissions simultaneously for these kinds of sources.

Several TDEs exhibit oscillations in their X-ray light curves, with some occurring on timescales of days. A notable example is AT2020ocn, which shows quasi-periodic modulations in its X-ray flux with a period of roughly 15 days (\cite{AT2020ocn_Pasham2024Natur}). These modulations possibly originate from the radiation pressure instability. Application of our model to this source faces the challenge of reducing the oscillation amplitude, as mentioned above, and requires further improvement of the model to partially suppress the instability. 
In addition, there are some fast oscillations occurring on timescales of hours observed in TDEs, which are typically recognized as quasi-periodic eruptions (QPEs) (\cite{Miniutti2023A&A_GSN069},\cite{Nicholl2024Natur}). The periods of QPEs are too short compared with the characteristic timescale of the radiation pressure instability in accretion disks in TDEs. To model the oscillations in QPEs with disk instability, one should set a local, narrow unstable zone that is injected with a low mass supply rate, as in \cite{Pan2022ApJ}. This is hard to reproduce with our global simulation of an accretion disk in TDE environment, i.e., injecting matter at $R_{\rm{c}}$ as equation (\ref{Rc}) in the form of $\dot{M}_{\rm{fb}}$ as equation (\ref{M_dot_fb}). Therefore, we do not attempt to explain QPEs with our disk model for now. Instead, we anticipate its potential application to oscillations with longer timescales that may be discovered in the future.

\section{Conclusions}
In this paper, we performed calculations of a 1D time-dependent accretion disk in the standard environment of TDEs, i.e., we inject materials at the circularization radius of the stellar debris in the form of $\dot{M}_{\rm{injct}}\propto t^{-5/3}$. 
The main results of our calculations are as follows:

\begin{itemize}
    \item The radiation pressure instability occurs when $\dot{M}_{\rm{fb}}$ evolves to a value close to $\dot{M}_{\rm{Edd}}$, leading to a sudden decline in luminosity by more than one order of magnitude.
    \item In many cases, the light curves then oscillate significantly. The oscillations periods become shorter for larger $\alpha$ and smaller $R_{\rm{out}}$. This could potentially explain the quasi-periodic flares observed on timescales of several days in some TDEs.
    \item When $\alpha$ is small or $R_{\rm{out}}$ is large, the oscillations in the light curves are completely suppressed. The light curves become flat after the steep decay induced by the radiation pressure instability.
\end{itemize}
Based on these calculations, we present an application of the light curves that become flat in the late times to the UV observations of ASASSN-15oi and ASASSN-14ae, together with the assumption of a photosphere. Our main results in this application are as follows:
\begin{itemize}
    \item The early-time steep decay followed by a late-time flattening feature in the UV light curves of these two sources could be a result of radiation pressure instability. 
    \item The photosphere radius obtained from fitting our model to the observations is much larger than the disk radius in the early times, then recedes to close to the disk size in the late times. This could indicate strong reprocessing in the early times. And in the late times, the UV flux of the photosphere model is similar to that of the disk model.
\end{itemize}
To account for other UV detections in TDEs with higher late-time luminosity, the radiation pressure instability should be partially suppressed to reduce the drop in the light curves, which can be achieved by including other factors such as modified viscosity, magnetic pressure and fallback heating, etc. 

So far we have only applied our model to the UV observations. Our accretion disk model has potential applications in explaining the soft X-ray emissions in TDEs, which are generally expected to come from the accretion disk. When the radiation pressure instability occurs, $T_{\rm{eff}}$ in the inner disk typically changes about an order of magnitude. We therefore expect to apply the disk instability model to X-ray light curves that show large-scale variability. This study will be carried out in detail in future work.

\section{Acknowledgments}
Chenlei Guo thanks Mingjun Liu for his very useful advice on the derivation of equation (\ref{energy conserv2}) and on numerical methods. We thank the referee for the very helpful comments, which have made the discussions of our main results more complete. We also thank the organizers and participants of the TDE Full Process Simulation Seminar Series for the valuable lectures and discussions. This work is supported by the Strategic Priority Research Program of the Chinese 
Academy of Science (Grant No. XDB0550200),
National Key R\&D Program of China (No. 2023YFA1607903) and National Natural Science Foundation of China (Grant No. 12333004, 12173048).

\appendix

\renewcommand{\thefigure}{\thesection\arabic{figure}} 
\setcounter{figure}{0}  

\section{Vertical integration of the disk structure}
\label{app:vertical}
The hydrostatic equilibrium in z direction is 
\begin{equation}
    \frac{dp}{dz}=-\rho \Omega^{2}z. \label{hydrostatic equilibrium dz}
\end{equation}
Following \cite{Hoshi1981}, we assume a polytropic relation in the vertical direction
\begin{equation}
    p=K\rho^{1+1/N}, \label{polytropic}
\end{equation}
where N is the polytropic index of the accretion disk. We take $N=2$ in this work which indicates an adiabatic index $\gamma=3/2$ between 4/3 and 5/3. The boundary condition is $p=0, \rho=0$ at $z=H$ and $p=p_{\rm{e}}, \rho=\rho_{\rm{e}}$ at $z=0$. The subscript $e$ indicates values on the disk equatorial plane. Equation (\ref{hydrostatic equilibrium dz}) can then be integrated vertically, which gives
\begin{equation}
    \rho=\rho_{\rm{e}}\left (1-\frac{z^{2}}{H^{2}} \right )^{2}, \label{rho_z}
\end{equation}
\begin{equation}
    p=p_{\rm{e}}\left (1-\frac{z^{2}}{H^{2}} \right )^{3}.
\end{equation}
The integration of (\ref{rho_z}) gives
\begin{equation}
    \Sigma = 2\int_{0}^{H}\rho dz=2\rho_{\rm{e}}\int_{0}^{H} \left (1-\frac{z^{2}}{H^{2}} \right )^{2} dz=\frac{16}{15}\rho_{\rm{e}}H,
\end{equation}
thus we can take $\Sigma\approx \rho_{\rm{e}}H$.

The integration of the total pressure $p$ gives
\begin{equation}
    \Pi = 2\int_{0}^{H}p dz=2p_{\rm{e}}\int_{0}^{H} \left (1-\frac{z^{2}}{H^{2}} \right )^{3} dz=\frac{32}{35}p_{\rm{e}}H.  \label{Pi}
\end{equation}

The integration of  equation (\ref{hydrostatic equilibrium dz}) also gives
\begin{equation}
    2(1+N)\frac{p_{\rm{e}}}{\rho_{\rm{e}}} = \Omega_{\rm{K}}^{2}H^{2}. \label{hydrostatic equilibrium N}
\end{equation}

Following \cite{Janiuk2002ApJ}, we define three coefficients $C_{1}, C_{2}$ and $C_{3}$ to better describe the vertically integrated disk structure, which are expressed as
\begin{equation}
    C_{1} = \frac{1}{p_{e}H}\int_{0}^{H}pdz ,
\end{equation}
\begin{equation}
    C_{2}= \frac{Q_{\rm{rad}}^{-}}{2}\frac{3\kappa \rho H}{4\sigma T^{4}},
\end{equation}
and
\begin{equation}
   C_{3}=\frac{p_{\rm{e}}}{\rho_{\rm{e}}}\frac{1}{\Omega_{\rm{K}}^{2}H^{2}}.
\end{equation}
According to (\ref{Pi}) and (\ref{hydrostatic equilibrium N}), we have $C_{1}=0.46$ and $C_{3}=1/6=0.17$. To calculate $C_{2}$, we evaluate the temperature gradient in the radiation flux $F(z)=-\frac{ac}{3\kappa \rho }\frac{dT^{4}}{dz}$ with $dp/dz$. For a disk that is radiation pressure dominated, it gives $F(z)\approx\frac{c}{\kappa}\Omega^{2}z$.  Thus 
\begin{equation}
    C_{2}=\frac{c}{\kappa}\Omega^{2}H \frac{3\kappa \rho_{\rm{e}} H}{4\sigma T_{e}^{4}}=\Omega^{2}H^{2}\frac{\rho_{\rm{e}}}{p_{\rm{e}}}=2(1+N)=6.
\end{equation}

\section{The steady disk solutions}
\label{app:steady solutions}
We obtain the stable solution of the accretion disk by solving the local energy balance equation.
For a steady disk, taking $\partial S/\partial t = 0$, the energy loss rate through horizontal advection can be expressed as 
\begin{equation}
    Q^{-}_{\rm{adv}} = \Sigma Tv_{\rm{r}}\frac{d S}{dr}=\frac{\dot{M}}{2\pi r^{2}}\frac{p}{\rho}\xi, \label{Q_adv_steady}
\end{equation}
where $\xi$ is assumed to be a constant. We take $\xi=1.5$ throughout this work following \cite{Watarai2006ApJ}. 
The energy conservation equation for a steady disk can then be written as
\begin{equation}
    2D(r)= 2C_{2}\frac{4\sigma T^{4}}{3\kappa \rho H}  + \frac{\dot{M}}{2\pi r^{2}}\frac{p}{\rho}\xi. \label{steady energy balance}
\end{equation}
where $D(r)$ is the viscous dissipation rate per unit disk face area in a steady disk, and can be expressed as, 
\begin{equation}
  D(r)=\frac{3GM_{\rm{BH}}\dot{M}(r)}{8\pi r^{3}}g(r). \label{D(r)}
\end{equation}
Here 
$g(r)=1-\sqrt{R_{\rm{ISCO}}/r}$.
Combining equation (\ref{Q_adv_steady}), (\ref{steady energy balance}) and (\ref{D(r)}) together with (\ref{total pressure}), (\ref{hydrostatic equilibrium}) and (\ref{Q_vis}), we obtain the solutions for the local steady disk with Newton's iteration method.

We demonstrate the solutions on the $\dot{m}-\Sigma$ plane at $5R_{\rm{S}}$, $20R_{\rm{S}}$, $50R_{\rm{S}}$ and $100R_{\rm{S}}$ as shown in Figure \ref{S_curve}, where $\dot{m}=\dot{M}/\dot{M}_{\rm{Edd}}$. We plot the S-shaped curves at different radii.
The S-shaped curve shifts to the upper right for larger disk radius. Consequently, the radiation pressure instability occurs earlier in disks with larger radius, as discussed in section \ref{effect of Rout}.

\begin{figure*}[ht!]
\centering
\includegraphics[scale=0.5]{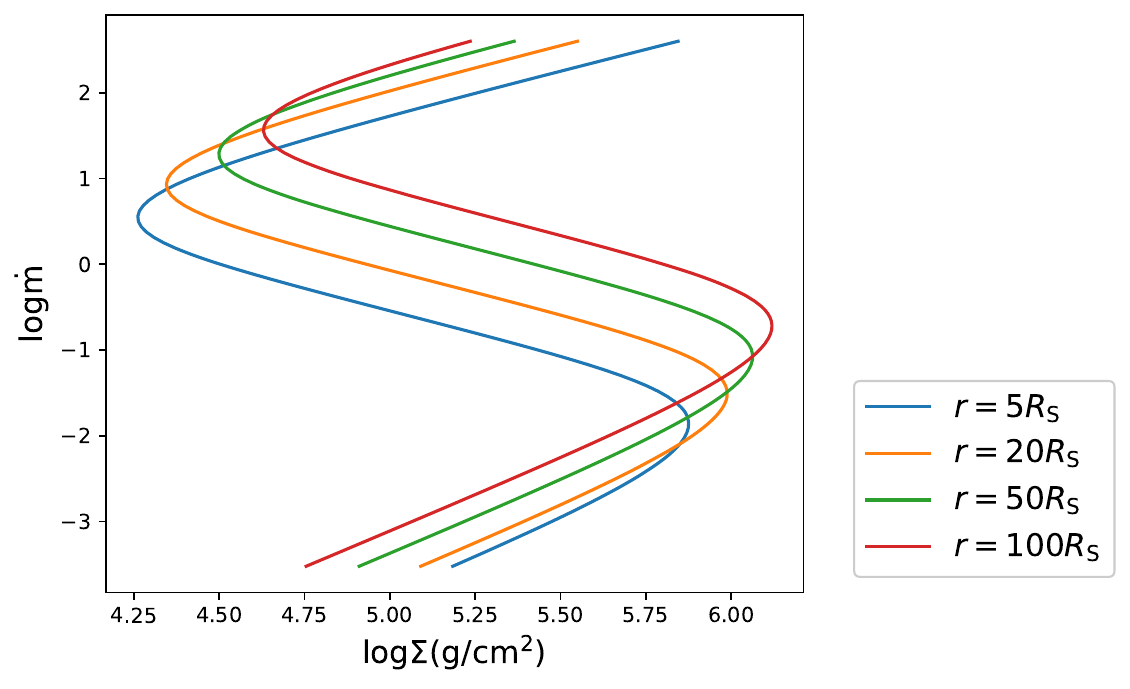}
\caption{Local stationary solution calculated with $M_{BH}=10^{6}M_{\odot}$ and $\alpha=0.01$, at $5R_{\rm{S}}$, $20R_{\rm{S}}$, $50R_{\rm{S}}$ and $100R_{\rm{S}}$. }
\label{S_curve}
\end{figure*}

\section{UV flux from the disk in application to observation }
\label{app:disk flux}

\begin{figure*}[ht!]
\centering
\includegraphics[scale=0.5]{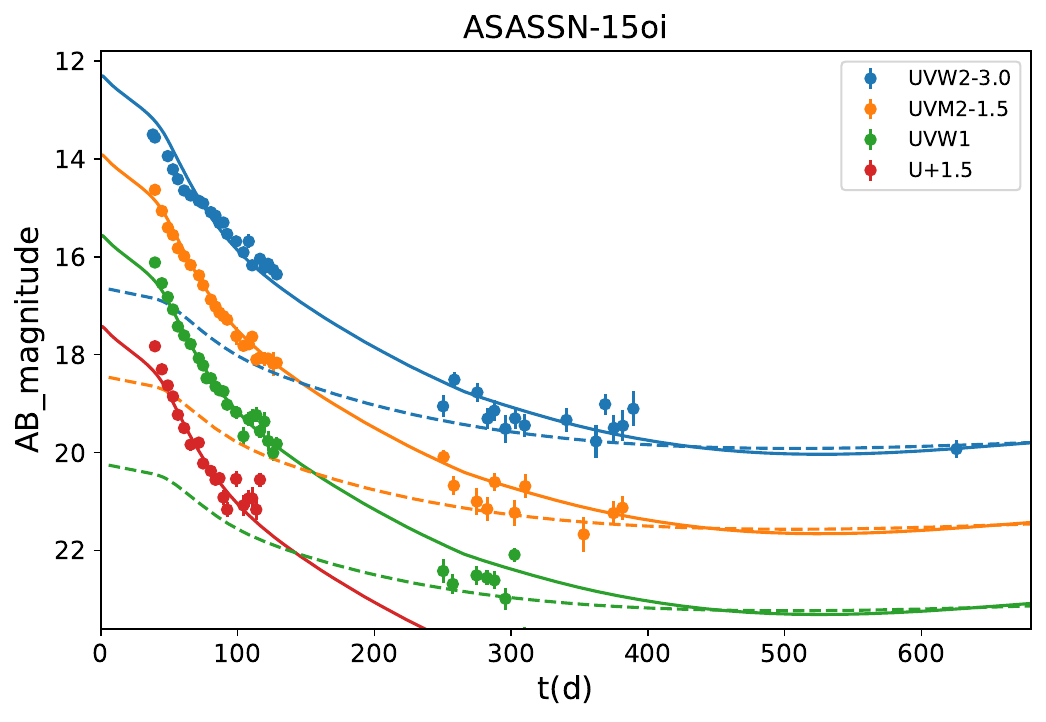}
\caption{Comparison between the UV light curves of the photosphere and that of the disk from our calculation for ASASSN-15oi. The solid lines are the UV flux of the photosphere, identical to the $L_{\rm{c}}/L_{\rm{Edd}}=0.003$ case in Figure \ref{15oi_lc_Rph}. The dashed lines show the UV flux of the disk.}
\label{15oi_disk}
\end{figure*}

\begin{figure*}[ht!]
\centering
\includegraphics[scale=0.5]{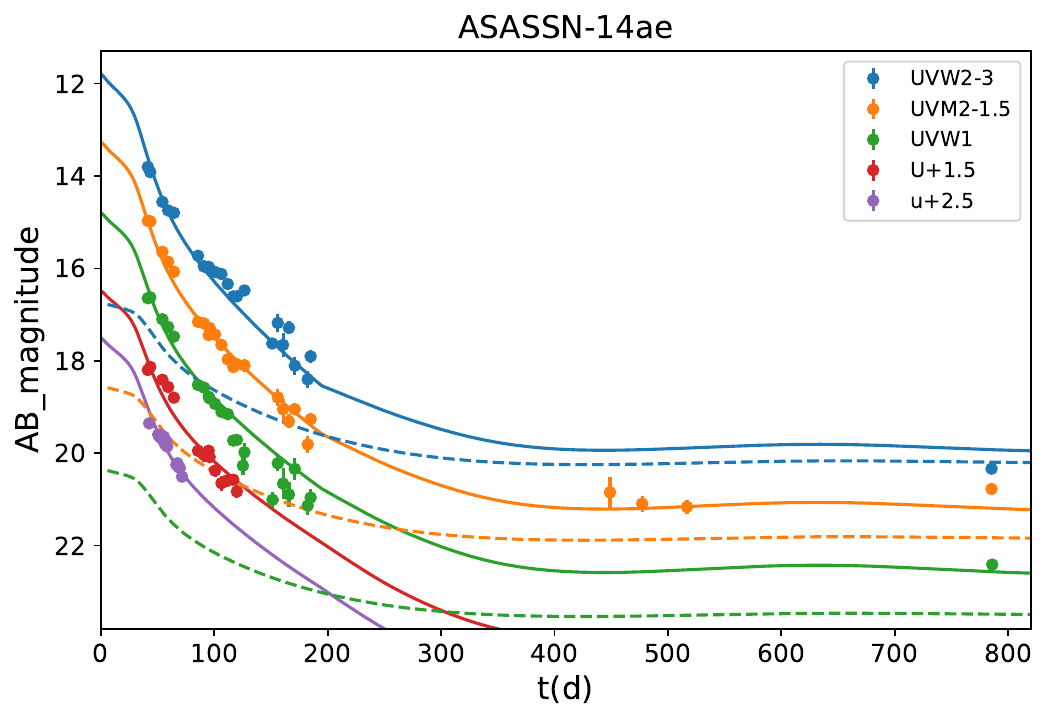}
\caption{Comparison between the UV light curves of the photosphere and that of the disk from our calculation for ASASSN-14ae. The solid lines are the UV flux of the photosphere, identical to the $L_{\rm{c}}/L_{\rm{Edd}}=0.003$ case in Figure \ref{14ae_lc_Rph}. The dashed lines show the UV flux of the disk.}
\label{14ae_disk}
\end{figure*}


\bibliography{sample7}{}
\bibliographystyle{aasjournalv7}



\end{document}